\documentclass[5p]{elsarticle}
\newcommand\lw{0.95}

\usepackage{lineno,amsmath}
\usepackage{amssymb}
\usepackage[colorlinks]{hyperref}
\modulolinenumbers[5]

\usepackage{url}
\usepackage{lipsum}
\usepackage{multicol}
\usepackage{multirow}

\usepackage{algorithm}
\usepackage{algpseudocode}

\newcommand\Tstrut{\rule{0pt}{2.6ex}}         
\newcommand\TstrutMini{\rule{0pt}{1.9ex}}         
\newcommand\Bstrut{\rule[-0.9ex]{0pt}{0pt}}   
\newcommand{\tit}[1]{\textit{#1}}
\newcommand{\ttt}[1]{\texttt{#1}}
\newcommand{\tbf}[1]{\textbf{#1}}
\newcommand{\trm}[1]{\textrm{#1}}
\newcommand{\mc}[3]{\multicolumn{#1}{#2}{#3}}
\newcommand{\mr}[3]{\multirow{#1}{#2}{#3}}

\journal{Computer Physics Communications}

\begin{document}

\begin{frontmatter}

\title{GPU-accelerated event reconstruction for the COMET Phase-I experiment}

\author[1]{Beomki Yeo\corref{cor1}} 
\ead{byeo@kaist.ac.kr}
\author[2]{MyeongJae Lee} 
\author[3]{Yoshitaka Kuno} 

\cortext[cor1]{Corresponding author}
\address[1]{Department of Physics, Korea Advanced Institute of Science and Technology (KAIST), Daejeon 34141, Republic of Korea}
\address[2]{Center for Axion and Precision Physics Research,
Institute for Basic Science (IBS), Daejeon 34051, Republic of Korea}
\address[3]{Department of Physics, Graduate School of Science,
Osaka University, Toyonaka, Osaka 560-0043, Japan}





\begin{abstract}
This paper discusses a parallelized event reconstruction of the COMET Phase-I experiment. The experiment aims to discover charged lepton flavor violation by observing 104.97 MeV electrons from neutrinoless muon-to-electron conversion in muonic atoms. The event reconstruction of electrons with multiple helix turns is a challenging problem because hit-to-turn classification requires a high computation cost. The introduced algorithm finds an optimal seed of position and momentum for each turn partition by investigating the residual sum of squares based on distance-of-closest-approach (DCA) between hits and a track extrapolated from the seed. Hits with DCA less than a cutoff value are classified for the turn represented by the seed. The classification performance was optimized by tuning the cutoff value and refining the set of classified hits. The workload was parallelized over the seeds and the hits by defining two GPU kernels, which record track parameters extrapolated from the seeds and finds the DCAs of hits, respectively. A reasonable efficiency and momentum resolution was obtained for a wide momentum region which covers both signal and background electrons. The event reconstruction results from the CPU and GPU were identical to each other. The benchmarked GPUs had an order of magnitude of speedup over a CPU with 16 cores while the exact speed gains varied depending on their architectures.

\end{abstract}


\end{frontmatter}

\section{Introduction}\label{sec:introduction}
\begin{figure}[!ht]
	\centering
	\includegraphics[width=\lw\linewidth]{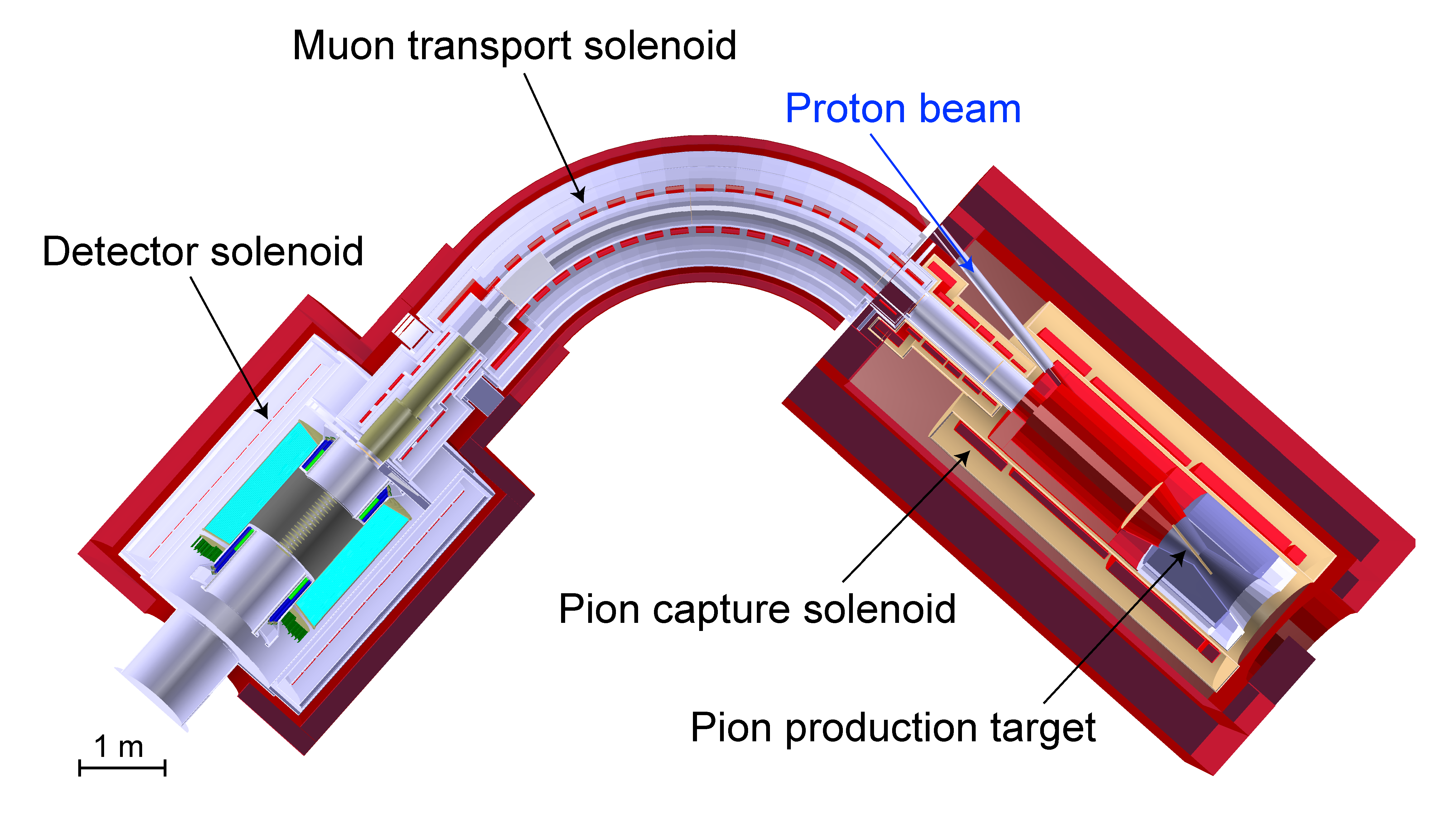}
	\caption{Layout of the COMET Phase-I experiment}
	\label{fig:phase1}
	
	\vspace{0.3cm}
	
	\includegraphics[width=\lw\linewidth]{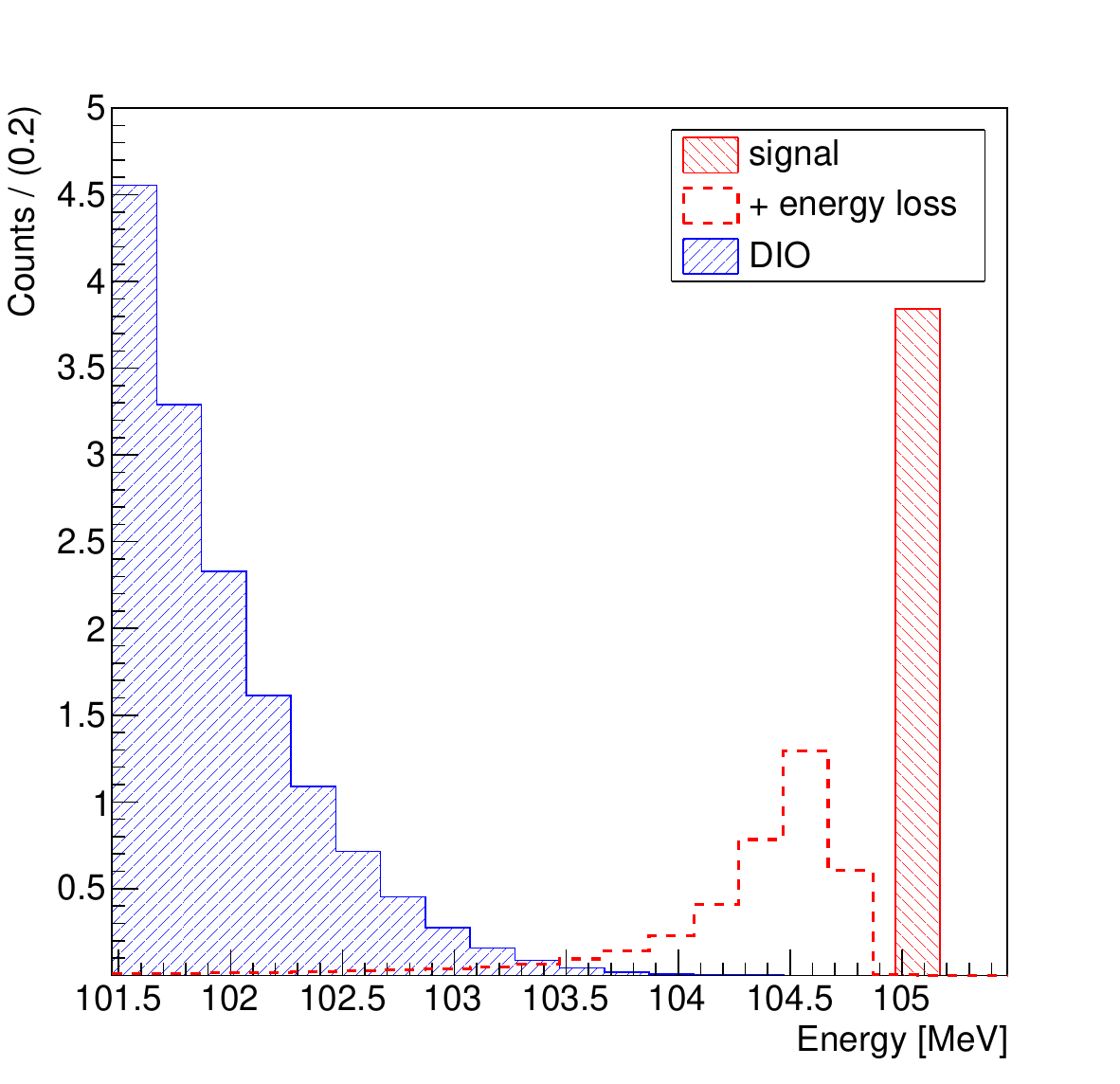}
	\caption{The red and blue histograms with the solid borderlines represent the theoretical initial energy of the signal and DIO electrons emitted from muonic atoms, respectively. The red histogram with the dashed borderline represents the signal electron energy distribution convoluted by a Landau distribution, corresponding to the energy loss inside the stopping targets. The Landau distribution was obtained by fitting the energy distribution of the simulated signal electrons that passed the muon stopping targets. It is assumed that the number of muonic atoms is $10^{15}$ and the branching ratio of $\mu^- +N \rightarrow e^- +N$ is $7\times 10^{-15}$.}
	\label{fig:DIOvsSignal}	
\end{figure}

The COMET experiment \cite{COMET_TDR}, located at J-PARC in Japan, will investigate charged lepton flavor violation by searching for neutrinoless muon-to-electron conversion in the field of a nucleus, namely $\mu^- + N(A,Z) \rightarrow e^- + N(A,Z)$. This transition is highly suppressed in the Standard Model with an unobservable branching ratio of $O(10^{-54})$ \cite{Petcov}. However, many extensions of the Standard Model predict that the transition rate may reach an observable level \cite{MuDecay}. In Phase-I of the COMET experiment, about $1.5 \times 10^{16}$ muons will be stopped in the target for 150 days of data acquisition. The expected upper limit of the muon-to-electron conversion is $7 \times 10^{-15}$ with a 90\% confidence level, which is about a 100 times improvement beyond the latest result \cite{SINDRUM2}. \\
\indent The schematic layout of the COMET Phase-I experiment is shown in Fig. \ref{fig:phase1}: A proton beam from the J-PARC main ring hits the graphite target to generate pions. The pions captured in the solenoid decay into muons, which are transported through the curved solenoid. Some fractions of the muons are stopped by the aluminum stopping target inside the detector solenoid and form muonic atoms. The muons stopped subsequently cascade down to the 1s orbital, followed either by muon captures by a nucleus (69\%) or muon decays (31\%). The neutrinoless muon-to-electron conversion can occur where the energy of signal electrons is monochromatic, of 104.97 MeV, which is equal to the muon mass subtracted by the 1s orbital binding energy and the nucleus recoil energy. Meanwhile, almost all muon decays  emit two additional neutrinos, i.e. $\bar{\nu_e}$ and $\nu_\mu$, as well as an electron to preserve the lepton flavor. This is called decay-in-orbit (DIO), and the emitted electrons have an endpoint energy close to 104.97 MeV, where the kinetic energies of two neutrinos are almost zero \cite{DIOspectrum}. Therefore, the detector should have a good momentum resolution to discriminate signal electrons from DIO electrons. It also should be noted that the actual energy of electrons we observe in the detector is reduced due to the energy loss inside the stopping targets. Figure \ref{fig:DIOvsSignal} shows how the energy loss affects the signal electrons.

\begin{figure}[!ht]	
    \includegraphics[width=\lw\linewidth]{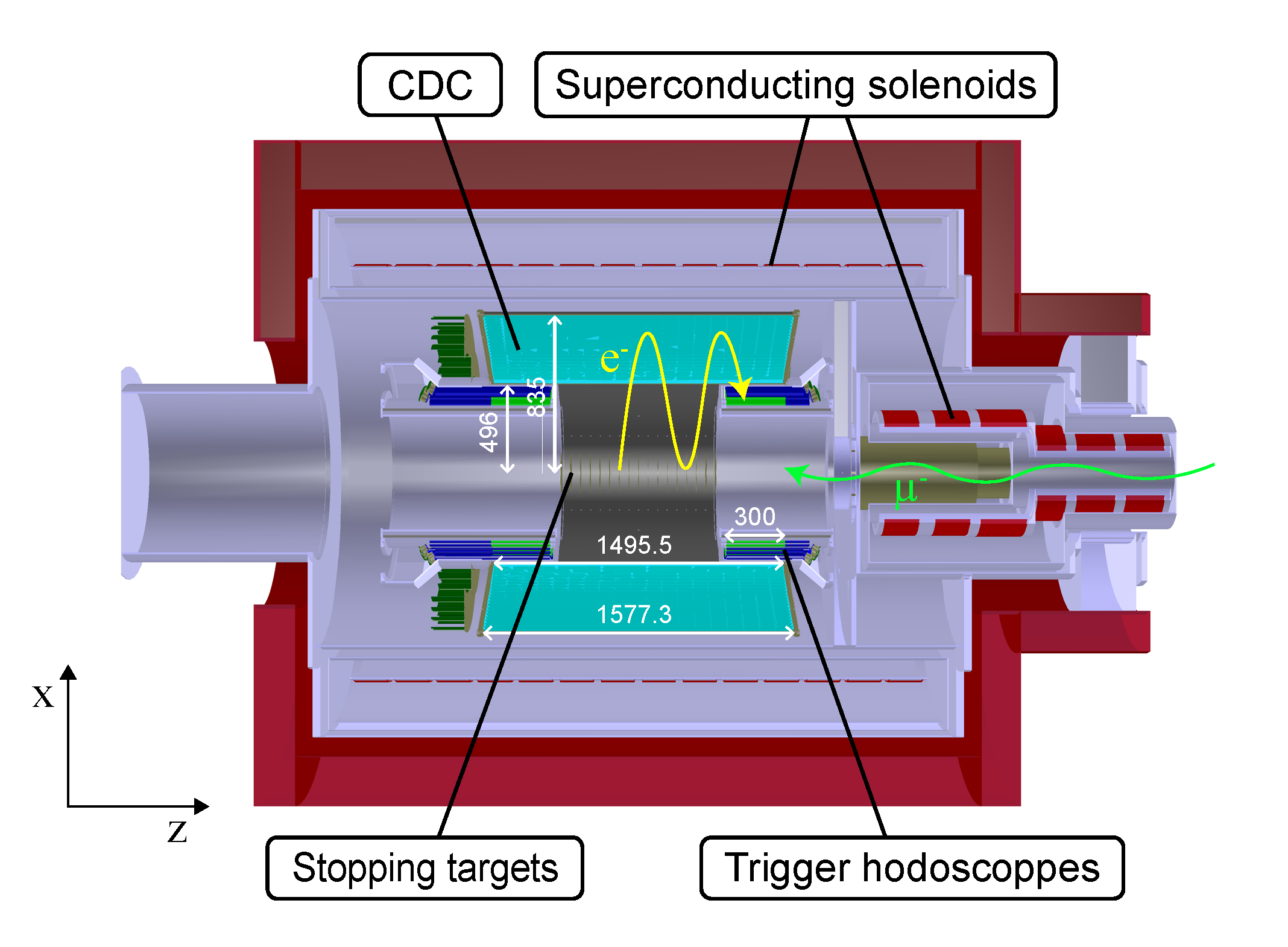}
    \caption{XZ cross section of the detector solenoid of the COMET Phase-I experiment. The yellow helix line shows a typical $e^-$ double turn event going toward the upstream direction. The sizes are in mm unit.}
    \label{fig:cdc}
\end{figure}

\begin{figure}[!ht]
\includegraphics[width=\lw\linewidth]{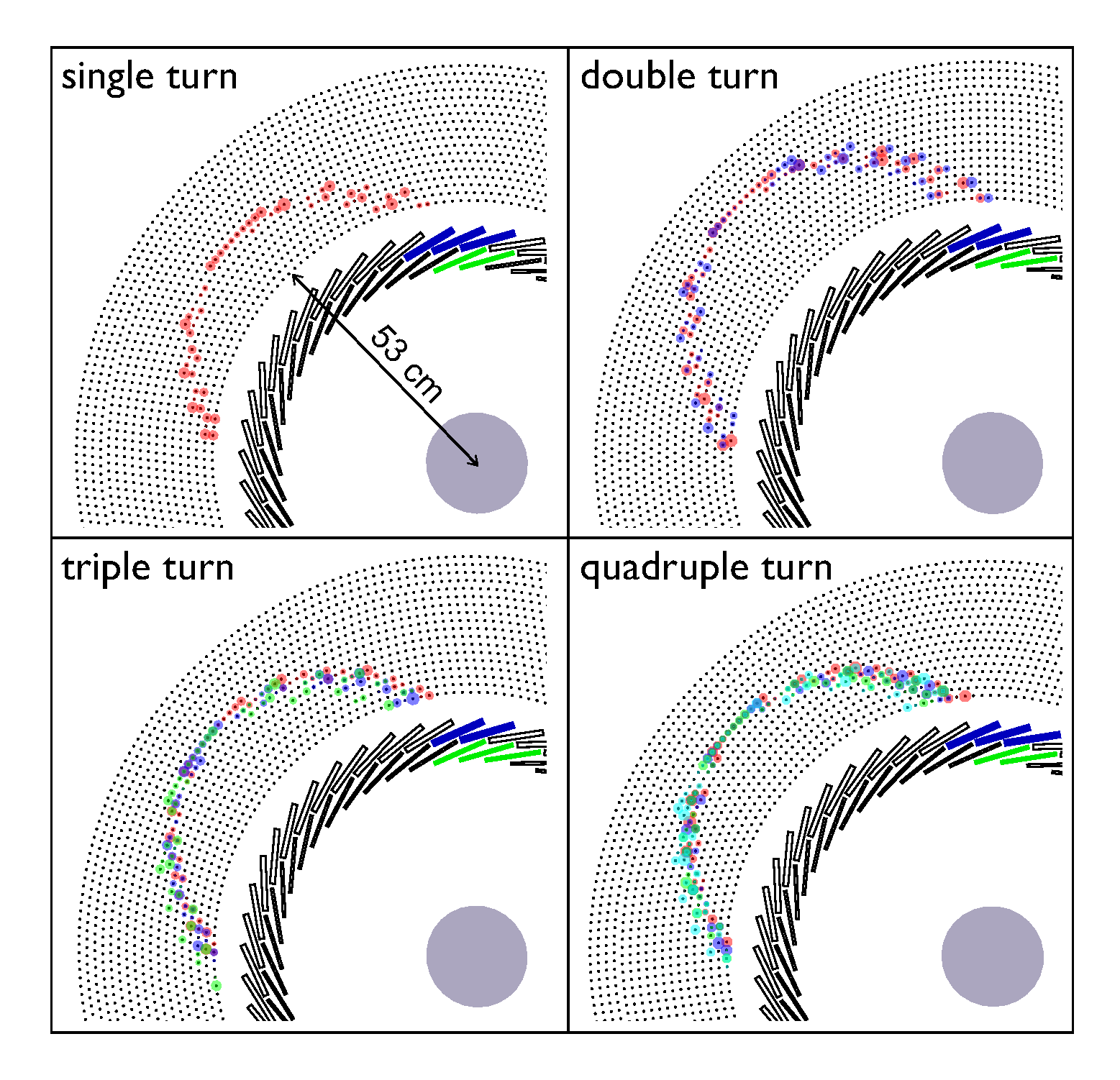}
\caption{XY cross section of the simulated events with various turn numbers. The transparent red, blue, green and cyan circles with small radii represent hits of the first, second, third and fourth turn partition, respectively. The radius of the circles is the drift distance of hit. The blue and green rectangles are triggered Cherenkov detectors and plastic scintillators, respectively. The gray circle with a 10 cm radius at the right-bottom represents the stopping targets.}
\label{fig:event_types}

\includegraphics[width=\lw\linewidth]{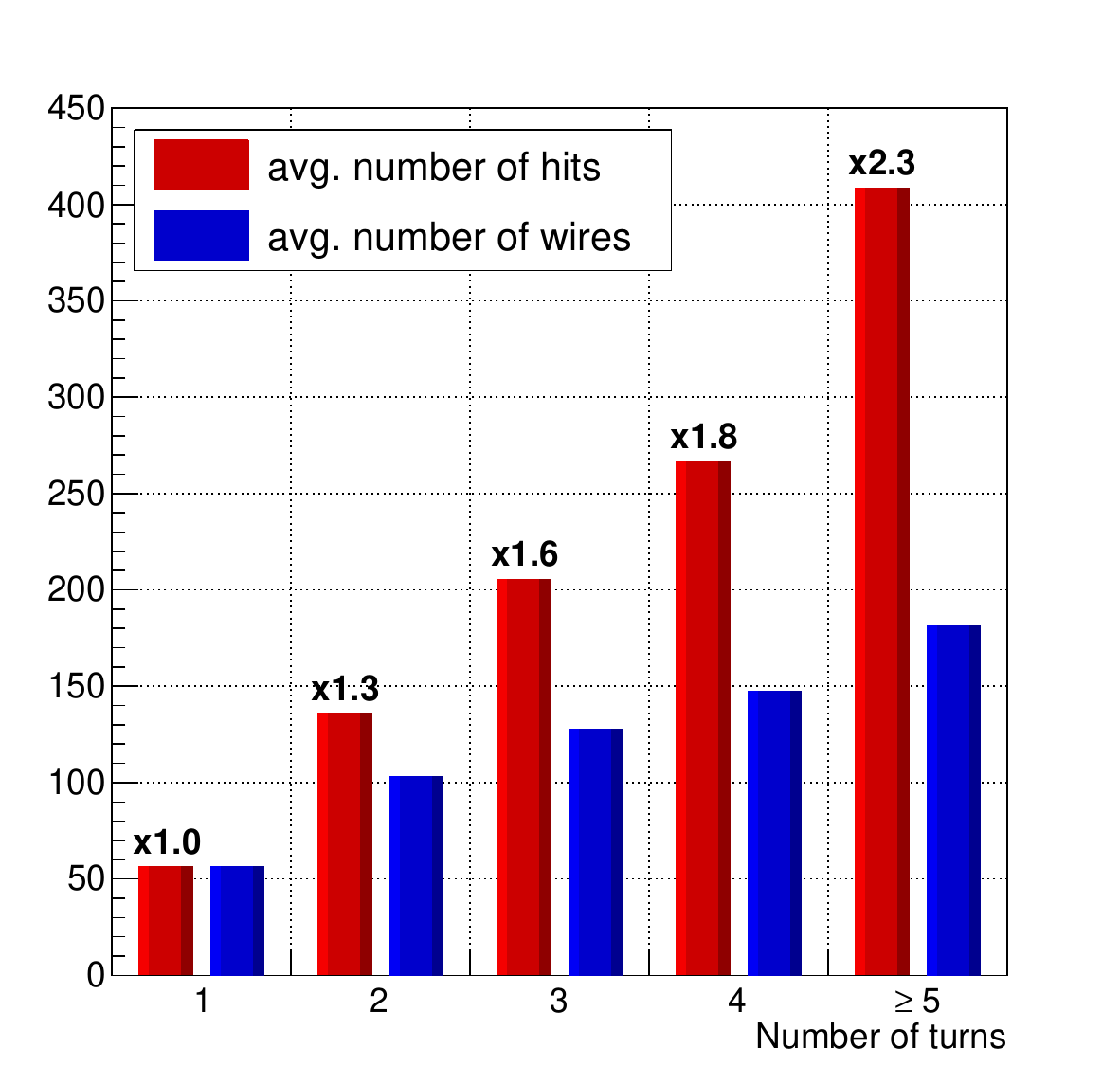}
\caption{The bar chart of the averaged number of hits (red) and wires that have hits (blue) from the simulated signal events. The numbers on the red bars represent the ratios between the two values, i.e., the average number of hits per wire.}
\label{fig:nHits_vs_nWires}
\end{figure}

\indent 
Inside the detector solenoid region (see Fig. \ref{fig:cdc}), the cylindrical drift chamber (CDC) is used as a detector to measure the momenta of electrons. Its inner and outer wall have radii of 49.6 cm and 83.5 cm, respectively, while the space in between is filled with a gas mixture of helium (He, 90\%) and isobutane (i-C$_4$H$_{10}$, 10\%) for the ionization of charged particles. A cluster of the ionized electrons, referred as a hit, drifts to the closest anode wire, and its drift distance $(r)$ is obtained by measuring the drift time. Now that an exact path relative to the wire is not provided explicitly, the so-called left-right ambiguity occurs with regard to whether the hit was generated on the left or right side of the wire. Hence, a ghost hit always persists at the opposite side of a real one, and they are referred as a left and right hit through the paper. \\
\indent 
Across 39 layers, there are 4986 anode and 14562 cathode wires made of gold-plated tungsten and pure aluminum, respectively. The radius of first layer is 51.4 cm and the gap between each layer is 0.8 cm. The 19 layers with even index are called field layers and consist of only the cathode wires. The 20 layers with odd index are called sense layers where the anode and cathode wires are placed alternatively. The first and last sense layers act as guard layers to remove the space charge that accumulates due to the ionization created in the regions between the CDC walls and guard layers. Therefore, the drift time is measured at the 18 sense layers without the guard layers. \\
\indent 
The wires are slightly tilted from the longitudinal axis by rotating them with respect to their radial vectors. The tilting angle is about 0.07 radian, and the tilting direction alternates between positive and negative signs for every layer. This configuration in stereo gives a resolution in the longitudinal position $(z)$ by calculating crossing points between the wires in adjacent layers. To obtain the momentum, the electron trajectory is reconstructed with a hypothetical track that connects the inferred positions of hits.\\
\indent At both the downstream and upstream ends of the CDC, trigger hodoscopes composed of 48 pairs of Cherenkov detectors and scintillators are positioned to identify electron events. Their fast timing response provides a reference time for the drift time measurement. Outside the outer wall, a superconducting solenoid applies a 1T magnetic field directed toward the downstream direction, which allows charged particles with momentum higher than 70 MeV/c to propagate in the CDC with a helical trajectory. \\
\indent Around 32\% of the signal electrons and 47 \% of the DIO electrons that satisfy the momentum cut of $>70$ MeV/c make multiple helix turns before they reach the trigger hodoscopes. The multiple turn events challenge the hit-to-turn classification because hits from different turns easily overlap in the same wire, as implied by Fig. \ref{fig:event_types}. The complexity of the multiple turn events increases with the turn numbers since the number of wires and the average number of hits per wire increase together, as shown in Fig. \ref{fig:nHits_vs_nWires}. This makes it hard to calculate $z$ position of a track due to the enormous number of possible combinations in matching the hits from the same turn partition.\\
\indent In this paper, we cope with the problem by introducing parallelization to accelerate the event reconstruction. The main idea is scanning the set of possible tracking seeds, where a seed represents the initial position and momentum of the track ($(\vec{x}, \vec{p})_0$). The goodness of the seed is evaluated by extrapolating a track from the seed in an inhomogeneous magnetic field and calculating the residual based on the distances between the hits and the extrapolated track. The hits close to the extrapolated track are classified as hits belonging to it. The scanning and extrapolating processes were fully parallelized on the NVIDIA GPU devices using CUDA (Compute Unified Device Architecture) which is a parallel computing platform \cite{CUDA}. \\
\indent The paper is organized as follows. Generalized parallelization schemes for a GPU and a multi-threaded CPU are described in Section \ref{sec:scheme}.Their applications to the simulated multiple turn events and tracking performance are described in Section \ref{sec:application}. Section \ref{sec:results} addresses technical aspects of the parallelization by benchmarking consumer and server-grade graphic cards. A discussion follows in Section \ref{sec:discussion}. 

\section{Parallelization scheme}\label{sec:scheme}

\subsection{Seed scanning and hit classification}\label{sec:seedscanning}
The residual sum of squares (RSS) for a seed is defined by:
\begin{equation}\label{eq:goodness}
\trm{RSS} = \sum _{i=1}^{N} \left[ \underset{j \in [1,m]}{\mathrm{min}} \{d^L_{ij},d^R_{ij},\lambda \} \right]^2 = \sum_{i=1}^{N} d^2_i,
\end{equation}
where $i$ and $j$ represent the index of the wire and hit, respectively. $N$ is the number of wires, and $m$ is the number of hits on the wire. $d^L_{ij}$ and $d^R_{ij}$ are the distance-of-closest-approach (DCA) of the left and right hits, respectively. DCA is obtained by finding the point-of-closest-approach (PCA) of the extrapolated track to the wire. $d_i$ is the minimum DCA bounded by $\lambda$, a cutoff value, which is introduced to ignore the contributions from outliers. The track extrapolation employs the Runge-Kutta-Nystr\"{o}m (RKN) method \cite{RKNMethod} with an adaptive step size that adjusts to the roughness of the magnetic field. \\
\indent The seeds to be scanned are selected with a detector-dependent prior knowledge of the possible ranges of each seed component. The method to obtain the ranges for the COMET Phase-I experiment is elaborated in Subsection \ref{subsection:apply_parallelization}. The calculation of the RSS is iterated over each seed by extrapolating a track with the RKN method. A grid search rather than a gradient descent method was used to find the global minimum since the RSS is a non-smooth convex function with respect to the seed components. 
\\ 
\indent The seed with the lowest RSS may not lead to the global minimum after the first iteration because there can be multiple local minima due to the event complexity. For that reason, several seeds with the lowest RSS are selected as reference points for the next seed scanning, where new seeds are fetched in their vicinity and with a finer granularity. To cover as many local minima as possible, the selected seeds are ensured to be sparse enough to avoid overlaps between new seed sets from each reference point. This seed selection process is repeated until sufficient resolution in the seed components is achieved.\\
\indent 
The finally selected seeds are used for hit classification: the closest hits of every wire whose $\trm{min}(d^L_{ij},d^R_{ij})$ is smaller than $\lambda$ are classified as the hits of the track. Since each finally selected seed may have a different set of classified hits, a precise fitting method based on Kalman filtering \cite{KALMAN} should be applied to them independently to evaluate and compare the qualities of the fitted tracks. \\
\indent Detailed implementations of the seed scanning for both a GPU and a multi-threaded CPU are explained in the next subsections.

\subsection{Implementation on GPU}

\begin{figure}[!h]
	\centering
	\includegraphics[width=0.7\linewidth]{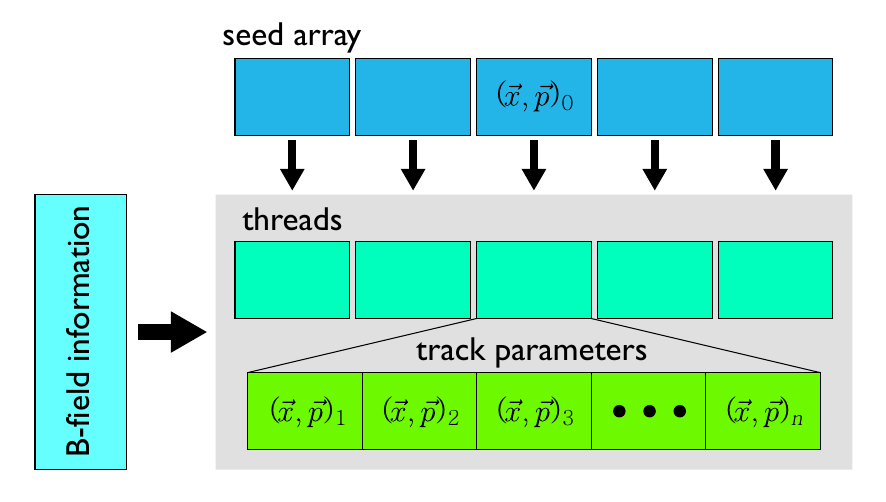}
	\caption{Schematics of the first kernel to record the track parameters for every step. GPU blocks are omitted in the schematics for better visibility.}
	\label{fig:first_kernel}
	
	\vspace{0.25cm}
	
	\includegraphics[width=\lw\linewidth]{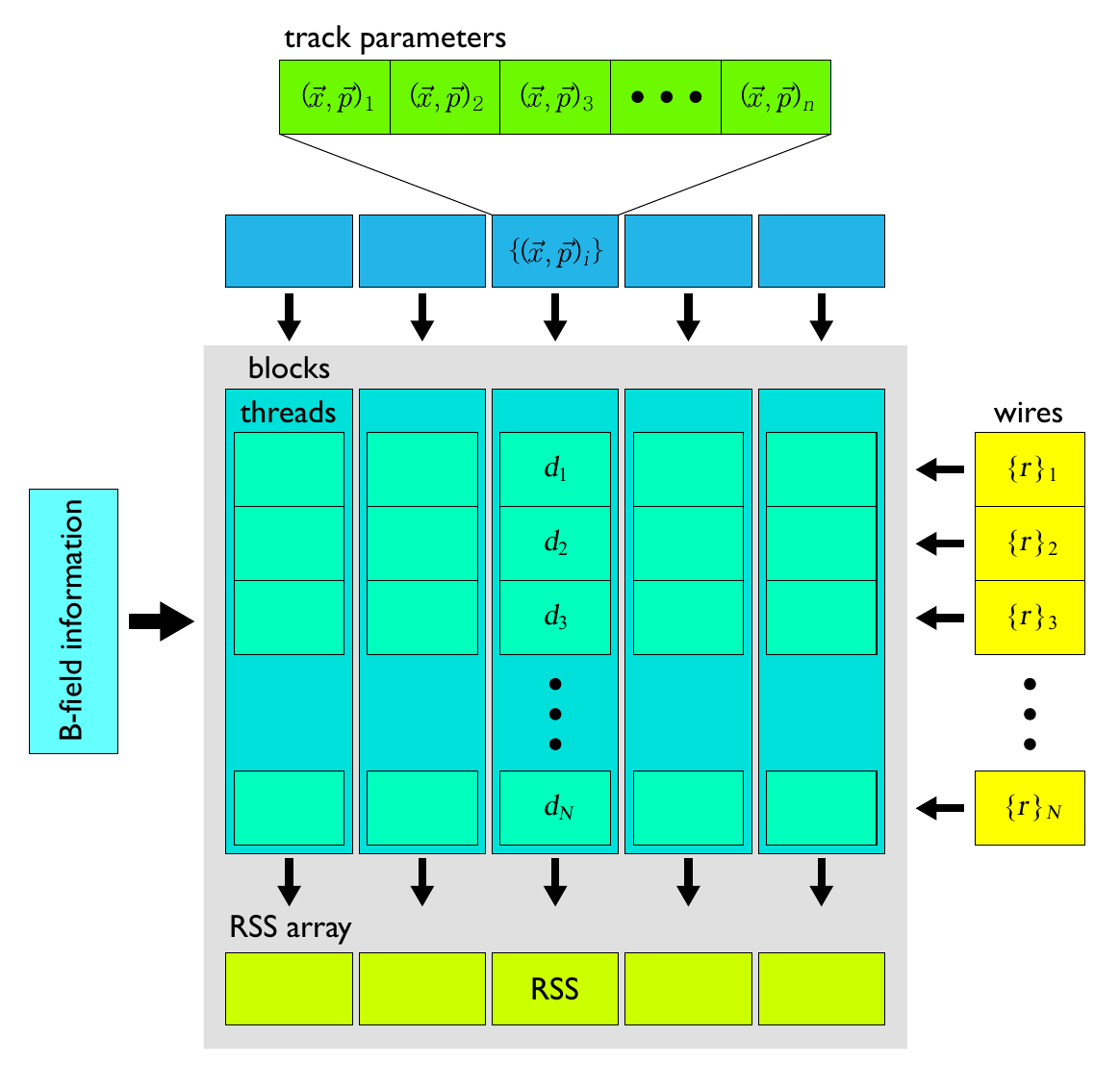}
	\caption{Schematics of the second kernel to find $d_i$ of each wire and obtain the RSS of each seed. $\{r\}_i$ represents the set of the drift distances of hits.}
	\label{fig:second_kernel}
	
	\vspace{0.25cm}
	
	\includegraphics[width=\lw\linewidth]{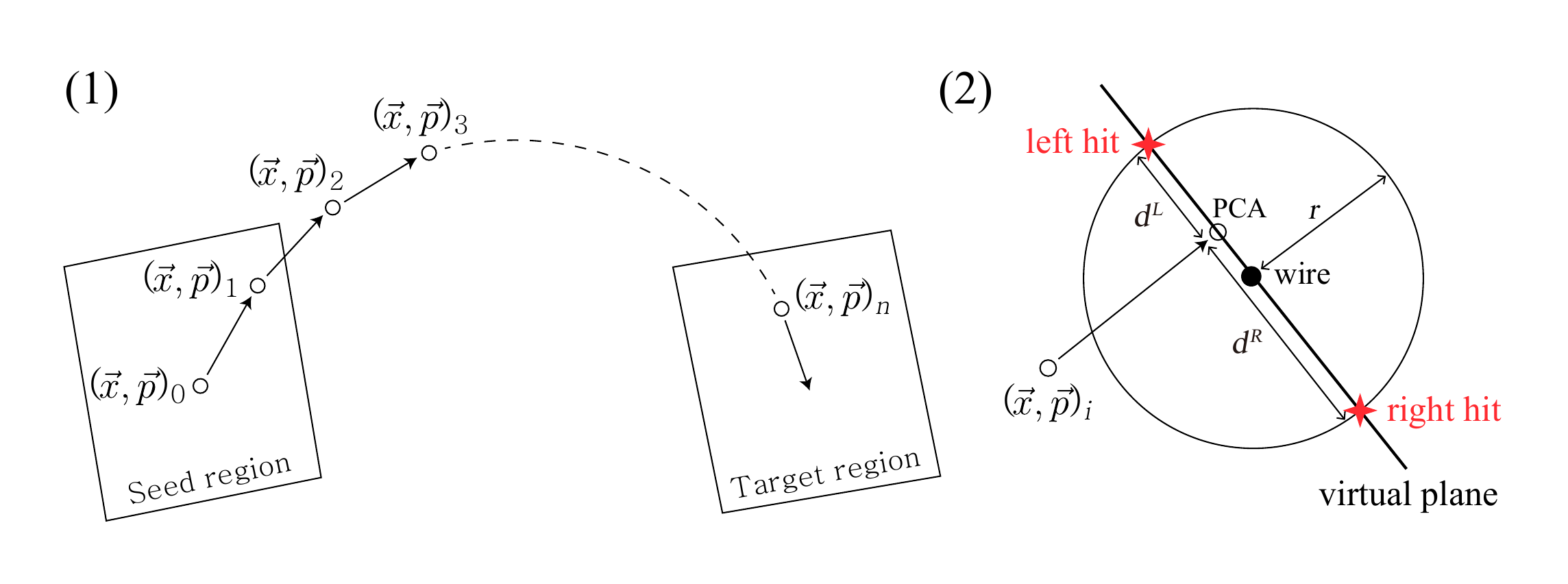}
	\caption{Illustrations of the tasks done per thread for each kernel function. (1) Track parameters recording and (2) DCA finding (view from wire axis).}
	\label{fig:illust}
	
	\vspace{0.25cm}
		
	\includegraphics[width=\lw\linewidth]{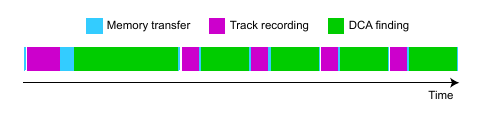}
	\caption{A typical timeline of the GPU processes during the seed scanning}
	\label{fig:gpu_timeline}
\end{figure}
\indent The architecture of a GPU is designed for parallelized computation by embedding thousands of cores. The resources of GPU devices are accessed by a CUDA kernel function that is executed with the aid of threads grouped into blocks. A block is composed of the hardware components called warps. The instructions on warps are scheduled by streaming multiprocessors (SMP) that constitute the whole GPU device. \\
\indent 
CUDA follows the SIMT (Single Instruction, Multiple Threads) execution model for parallel computing. Unlike SIMD (Single Instruction, Multiple Data), where common instructions are always executed simultaneously over vectorized data, the SIMT model allows  a conditional branch over vectorized threads in the same warp. However, instructions for different branches are only operated serially by disabling the threads not on the current branch. Such divergence in thread instructions is one of the major reasons for GPU performance degradation. \\
\indent The algorithm was divided into two kernel functions to minimize the divergence during the calculation of the RSS: (1) Track recording where track parameters, the position and momentum, on the extrapolated trajectory are recorded for every seed and (2) DCA finding where the track parameter closest to the wire serves as a new seed to find the DCA within a few extrapolation steps. Most of the computations in the kernels were based on double precision floating point operations.

\begin{algorithm*}[h!]
	\caption{A pseudocode for the CUDA kernel function: Track recording \TstrutMini}
	\label{alg:track_recording}
	\begin{algorithmic}[1] 
		\Function{TrackRecording}{\ttt{seeds}, \ttt{BFields}, \ttt{\&trackParametersSet}, $\ldots$}	
			\State {\ttt{gid}  $\gets$ \ttt{threadId+blockSize$\times$blockId}} \Comment{Global ID for the seeds}
			\State {\ttt{state} $\gets$ \ttt{seeds[gid]}}			
			
			\State {\ttt{nPars} $\gets$ 0}
			\While {\ttt{state} $\not\in$ \ttt{targetRegion}} \Comment{Loop until \ttt{state} reachs \ttt{targetRegion}}
				\State {Update \ttt{state} with an adaptive step size \Comment{RKN extrapolation with \ttt{BFields}}}
				\State {\ttt{trackParametersSet[gid][nPars]} $\gets$ \ttt{state}}
				\State {\ttt{nPars=nPars+1}}
			\EndWhile
		\EndFunction
	\end{algorithmic}
\end{algorithm*}

\begin{algorithm*}[h!]
	\caption{A pseudocode for the CUDA kernel function: DCA finding \TstrutMini}
	\label{alg:dca_finding}
	\begin{algorithmic}[1] 
		\Function{DCAFinding}{\ttt{trackParametersSet}, \ttt{wireGeoms}, \ttt{wireMeasurements}, \ttt{BFields}, \ttt{\&RSS}, $\ldots$}
		\State {\ttt{tid} $\gets$ \ttt{threadId} \Comment{Thread ID for the wires}}
		\State {\ttt{bid} $\gets$ \ttt{blockId}  \Comment{Block ID for the seeds}}
		\State {\ttt{geom $\gets$  wireGeoms[tid]} \Comment{Both endpoints of the wire}}
		\State {\ttt{meas $\gets$  wireMeasurements[tid]} \Comment{Set of the drift distances}}
		\State {\ttt{params $\gets$ trackParametersSet[bid]}}
		\State {Find \ttt{state ($\in$ params)} closest to \ttt{geom}}
		\While {true}
			\State {Define \ttt{virtualPlane} with \ttt{(state,geom)}}
			\State {Get \ttt{distance} between \ttt{state} and \ttt{virtualPlane}}
		    \If {\ttt{distance $<$ TOLERANCE} } \Comment{Break the loop if \ttt{state} gets close enough to \ttt{virtualPlane}}
				\State {\tbf{break}}
			\EndIf			
			\State {Update \ttt{state} with the fixed step size of \ttt{distance} \Comment{RKN extrapolation with \ttt{BFields}}}
		\EndWhile
		\State{Get \ttt{DCA} with \ttt{(state,meas,$\lambda$)} \Comment{\ttt{DCA} is $d_i$, and $\lambda$ is the cutoff value of Eq. (\ref{eq:goodness})} }
		\State{\ttt{RSS[bid] $\gets$ RSS[bid]+DCA$^2$}}
		
		\EndFunction
	\end{algorithmic}
\end{algorithm*}

\begin{algorithm*}[h!]
	\caption{A pseudocode for a CPU thread to obtain the RSS of a seed \TstrutMini}
	\label{alg:cpu_function}
	\begin{algorithmic}[1] 
		
		\Function{GetRSS}{\ttt{seed}, \ttt{wireGeoms}, \ttt{wireMeasurements}, \ttt{BFields}, \ttt{\&RSS}, $\ldots$}
		
		\State  Get \ttt{trackParameters} by recording a track from \ttt{seed}	\Comment{Corresponds to the while loop of Algorithm \ref{alg:track_recording}}
		\For{\ttt{(geom, meas)} $\in$ \ttt{(wireGeoms, wireMeasurements)}} 
		\State {Find \ttt{state ($\in$ trackParameters)} closest to \ttt{geom}}
		\State {Update \ttt{state} until it gets close enough to \ttt{virtualPlane}} \Comment{Corresponds to the while loop of Algorithm \ref{alg:dca_finding}}
		\State {Get \ttt{DCA} with \ttt{(state,meas,$\lambda$)}}
		\State {\ttt{RSS $\gets$ RSS + DCA$^2$} }
		\EndFor
		
		\EndFunction
	\end{algorithmic}
\end{algorithm*}

\subsubsection{Kernel function 1: Track recording}
The seed set and magnetic field information in the grid format are transferred into the GPU device memory. Each thread extrapolates a track from each seed to the target region: the track parameters $((\vec{x}, \vec{p})_i)$ of all steps are recorded and transferred to the host memory on the CPU side. The overall schematic of the kernel function is described in Figs. \ref{fig:first_kernel} and \ref{fig:illust} (left part), while its implementation as a pseudocode is detailed in Algorithm \ref{alg:track_recording}.

\subsubsection{Kernel function 2: DCA finding}
The recorded track parameters of all seeds are transferred to the GPU with the magnetic field information and the drift distance information of hits. 
Each block calculates the RSS of each seed, and each thread in a block finds $d_i$ of each wire by starting off an extrapolation from the track parameter closest to the wire. Therefore, the number of threads assigned in a block is the same as the number of wires. The extrapolation step size is not adaptive but determined to be the distance between the track position and a virtual plane: The virtual plane, which is updated for every step, is normal to the track momentum and contains the wire axis. The extrapolation repeats until the new track point gets close enough to the virtual plane. Once the DCAs are found by all threads, the RSS is calculated for every block and transferred to the host memory. Figures \ref{fig:second_kernel} and \ref{fig:illust} (right part) outline the schematics of the kernel function and Algorithm \ref{alg:dca_finding}  represents its implementation as a pseudocode.\\
\indent 
Figure \ref{fig:gpu_timeline} shows a typical timeline of the GPU processes executed during the seed scanning. The kernel execution time of the first iteration is longer than that of the next iterations because the number of seeds prepared initially is larger than the number of seeds collected from the reference points. The data transfer in the figure includes both host-to-device and device-to-host transfers, where they could overlap with the kernels to reduce the entire processing time. However, such an asynchronous process was not utilized because the data transfer time was too short compared to the kernel execution time.

\subsection{Implementation on CPU}
Compared to a GPU, a CPU has a faster core clock speed and a relatively large portion of cache while having fewer cores and less memory bandwidth. These different features make a CPU efficient in processing serial tasks with a low latency.  A function for CPU was thus rearranged to allocate more tasks to a thread: each thread executes both track recording and DCA finding to calculate the RSS of each seed. The pseudocode for this unified function is detailed in Algorithm \ref{alg:cpu_function}.

\section{Application and results}\label{sec:application}

\subsection{Application of the parallelization scheme} \label{subsection:apply_parallelization}
\begin{figure}[!h]
	\centering
	\includegraphics[width=\lw\linewidth]{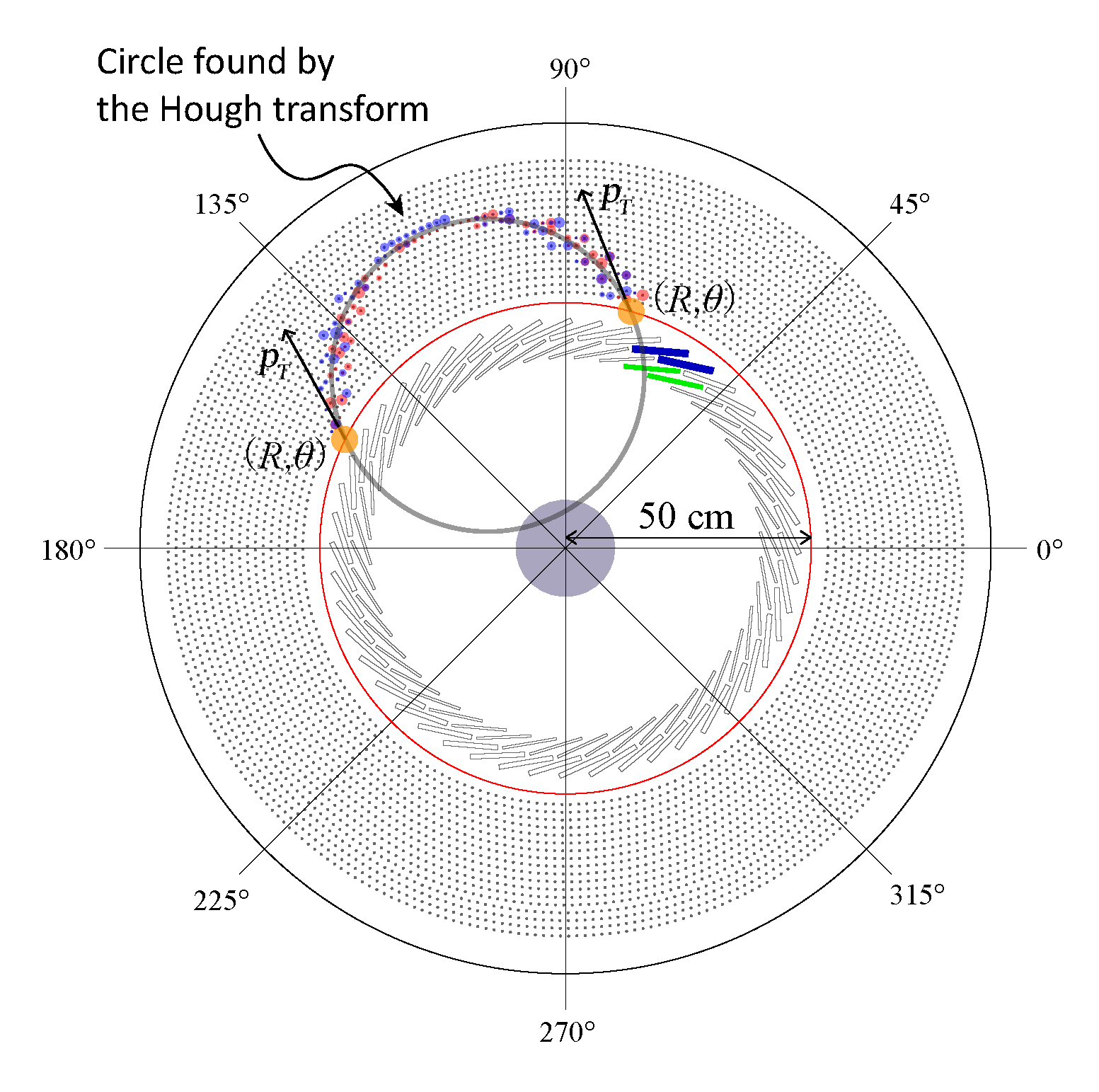}
	\caption{Scheme for extracting the transverse seeds $(\theta, p_x, p_y)$ from a circle found by the Hough transform. The seeds for the CDC entrance and exit are defined at the orange-shaded regions, where $R$ is fixed to 50 cm. The transverse momentum $(p_T)$ or $(\theta, p_x, p_y)$ is obtained from the radius and tangent line of the found circle.}
	\label{fig:transverse_seed}
	
	\vspace{0.3 cm}
	
	\centering
	\includegraphics[width=\lw\linewidth]{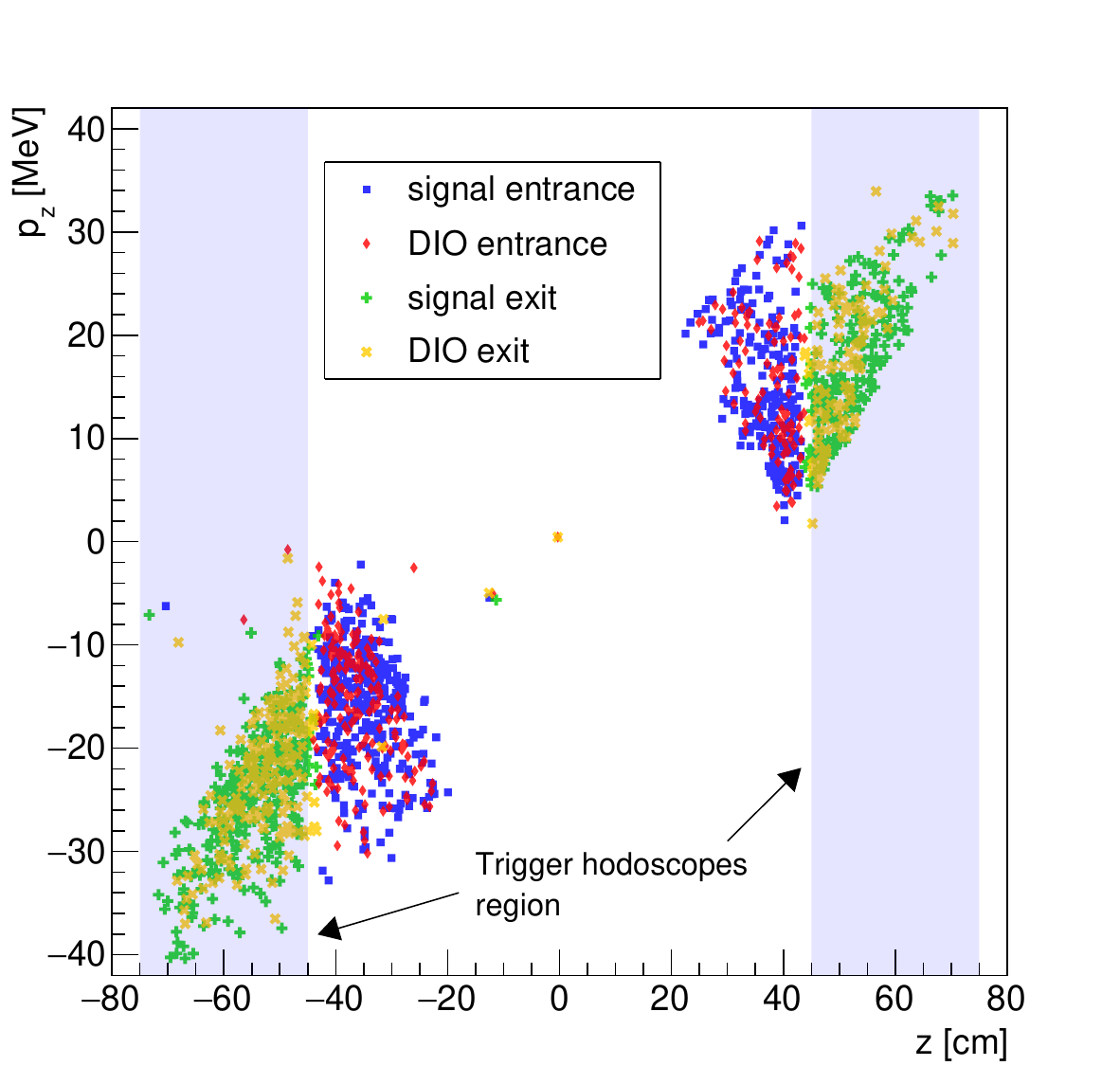}
	\caption{Longitudinal seed $(z,p_z)$ distributions of the last turn partition of the multiple turn events. The shaded areas at left and right side are where the trigger hodoscopes are positioned for the downstream and upstream directions, respectively.}
	\label{fig:zpz_distribution}
\end{figure}

\begin{figure}[!h]
	\centering
	\includegraphics[width=\lw\linewidth]{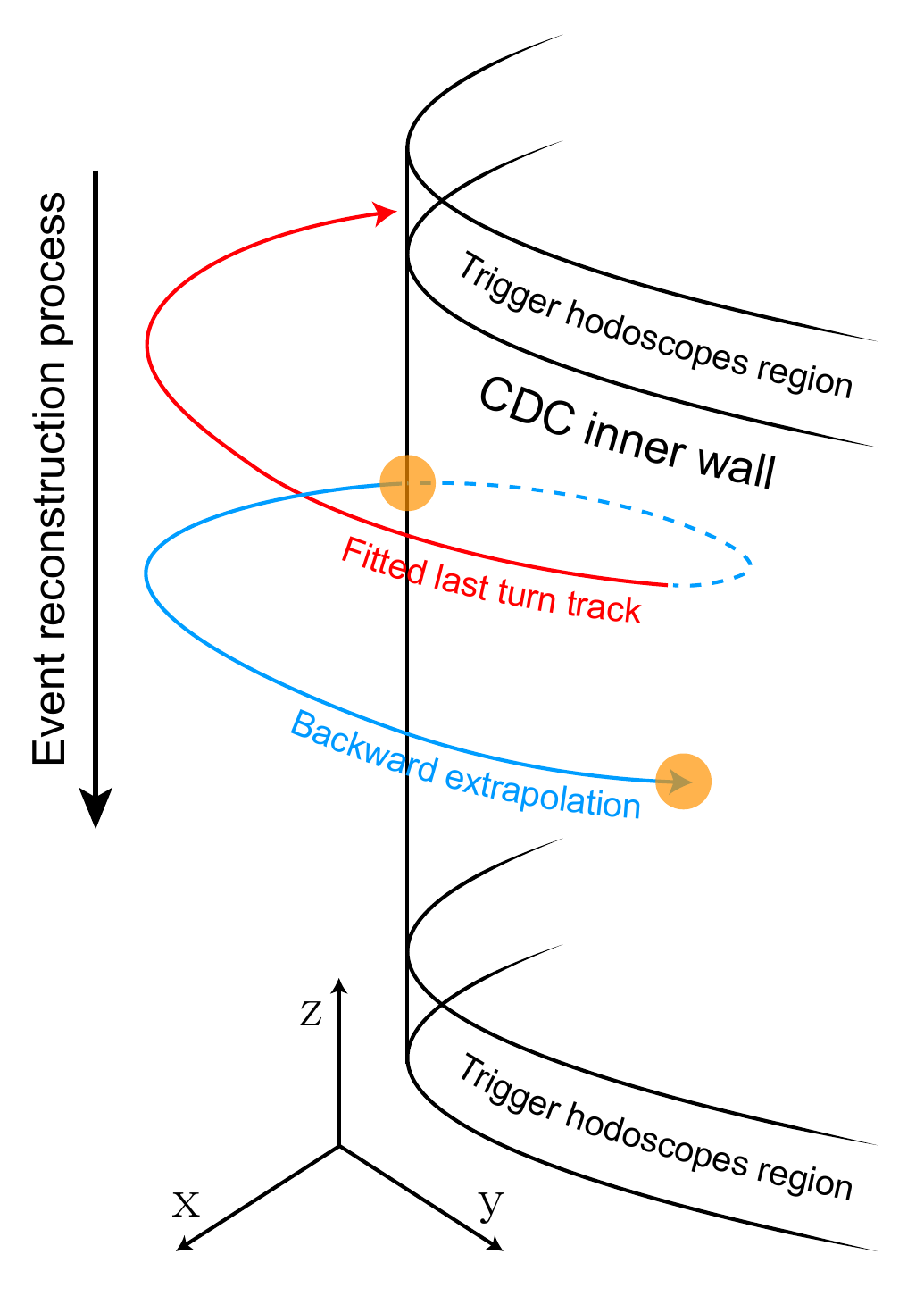}
	\caption{Backward extrapolation scheme to obtain the longitudinal seeds $(z,p_z)$ of the previous turn partition after fitting the last turn partition. The seeds for the CDC entrance and exit are collected at the orange-shaded regions, where $R$ is fixed to 50 cm.}
	\label{fig:backward_extrapolation}
\end{figure}

\begin{figure}[!h]
	\centering
	\includegraphics[width=\lw\linewidth]{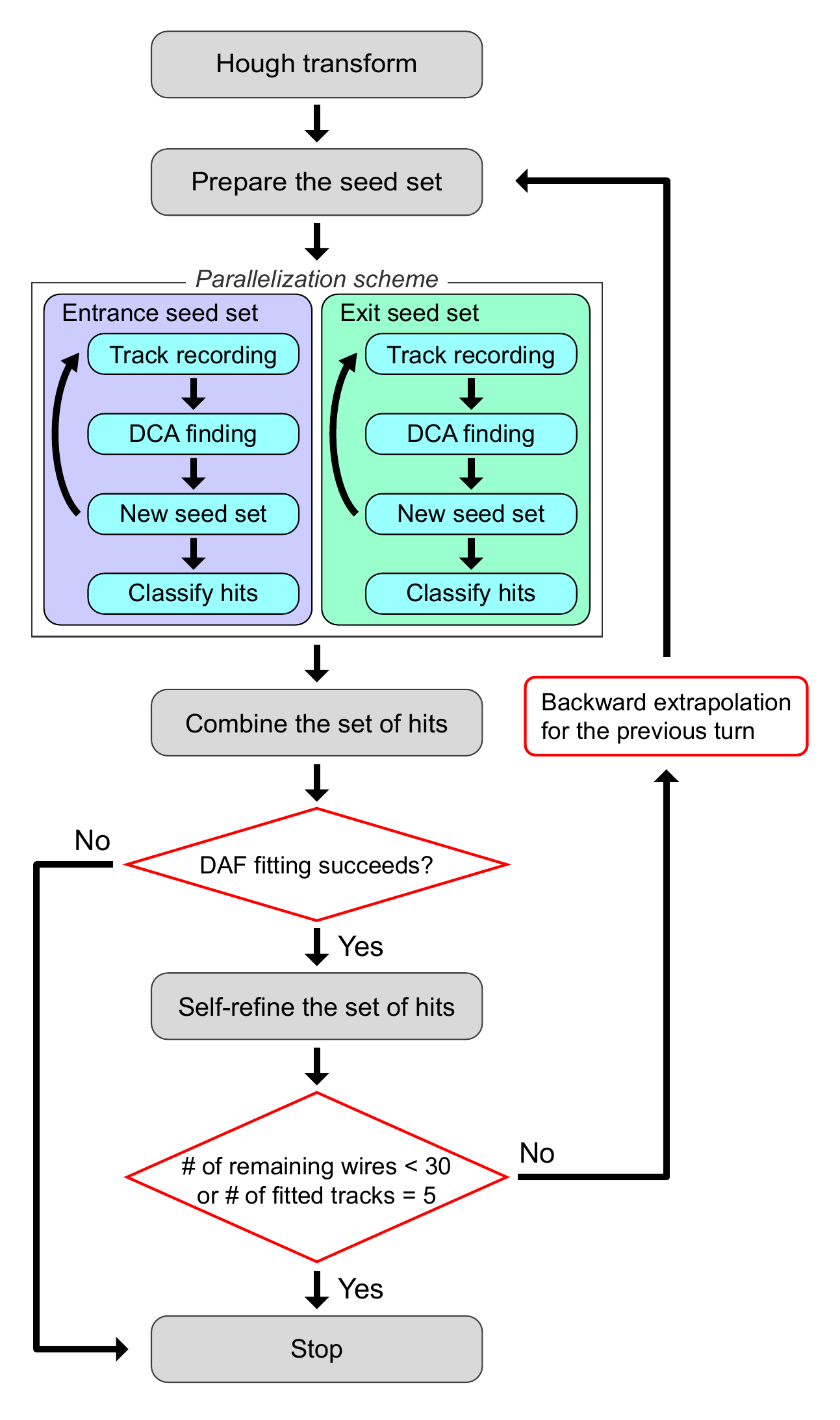}
	\caption{Flowchart for the event reconstruction process}
	\label{fig:process}
\end{figure}

The multiple scattering in material is the main issue to consider in applying the parallelization scheme to the multiple turn events. The scattering inside the detector materials induces uncertainty of the trajectory, and ambiguity of the track parameters increases with the total extrapolation length. Therefore, instead of fitting an event with a single continuous track, discrete tracks that are distant in the $z$ direction were fitted for each turn partition. In order to reduce the extrapolation length further, a turn partition was split into two half tracks whose seeds are at the CDC entrance and exit sides, respectively. The track extrapolation was done in the forward direction from the entrance and the backward direction from the exit by flipping the charge and momentum. The hits of the event were also grouped into two sets for each half track. \\
\indent For the both sides, the seeds were defined in front of the first layer of wires at the fixed radius of 50 cm, where we have five degrees of freedom for the seed: $(\theta, z)$ for the position in the cylindrical coordinate and $(p_x,p_y,p_z)$ for the momentum in the Cartesian coordinate. The transverse components of $(\theta, p_x, p_y)$ were extracted by finding a circle that is the transverse projection of a helical trajectory, as illustrated in Fig. \ref{fig:transverse_seed}. To find the circle, the Hough transform \cite{HoughTransform} was applied to the set of transverse positions of the fired wires. The resolutions of $(\theta, p_x, p_y)$ were estimated to be 0.02 radian, 3.2 MeV/c, and 3.2 MeV/c, respectively, for both the entrance and exit. \\
\indent The ranges of the longitudinal parameters $(z,p_z)$ of the last turn partition are also constrained because triggering electron trajectories end up at the trigger hodoscopes, as shown in Fig. \ref{fig:zpz_distribution}. Since all seed components of the last turn partition are predictable with known ranges, the event reconstruction starts from the end of a solenoid by fitting the last turn partition. A seed set was prepared by combining each component from the grids of the ranges. The number of seeds in the parallelization scheme was a few tens of thousands. \\
\indent Finally selected 10 seeds at each side have their own half sets of classified hits, while the seeds which have any missing layer without the hits were excluded. The complete sets of classified hits were formed by combinatorially merging two half sets, one of which is from each direction. Each complete set was fitted with a deterministic annealing filter (DAF) method \cite{DAEM,DAF} implemented in the GENFIT2 package \cite{GENFIT, GENFIT2}. The DAF method is an iterative Kalman filtering method that weights left and right hits on a scale from 0 to 1, where one with a low weight is regarded as an outlier. If the fitting results were not converged for any of the complete sets, the event reconstruction process was stopped.\\
\indent The quality of each complete set was evaluated by the number of degree of freedom (NDF) of the converged fitting results. The complete set with the highest NDF was self-refined by repeating the DAF method and the following hit selection processes. (1) A hit was removed from the set of classified hits if both its left and right hit had weights lower than 0.9. (2) If any hits in the wire were not classified, their $d^L_{ij}$ and $d^R_{ij}$ were obtained again by extrapolating a track from the fitted track parameter of the adjacent wire. We introduced a new cutoff value ($\eta$) to classify the closest hit whose $\trm{min}(d^L_{ij},d^R_{ij})$ is smaller than it.
\\
\indent  The fitted track was backward extrapolated to the seed regions where new longitudinal seeds were collected for the previous turn partition, as illustrated in Fig. \ref{fig:backward_extrapolation}. For the transverse seeds, the same seed set from the Hough transform was provided because the resolution of the Hough transform was still better than the extrapolated values. The hits classified previously were excluded in the seed scanning; therefore, the number of wires gets reduced whenever a track is fitted. The event reconstruction was continued until the number of remaining wires became less than 30 or the number of fitted tracks became five. A flowchart of the entire event reconstruction process is shown in Fig. \ref{fig:process}. To avoid confusion, we define the following terms: (1) an \tit{iteration} is a unit of the track recording and DCA finding and (2) a \tit{stage} represents the seed scanning for a turn, which includes several iterations. 

\begin{figure}[h!]
	\centering
	\includegraphics[width=0.808\linewidth]{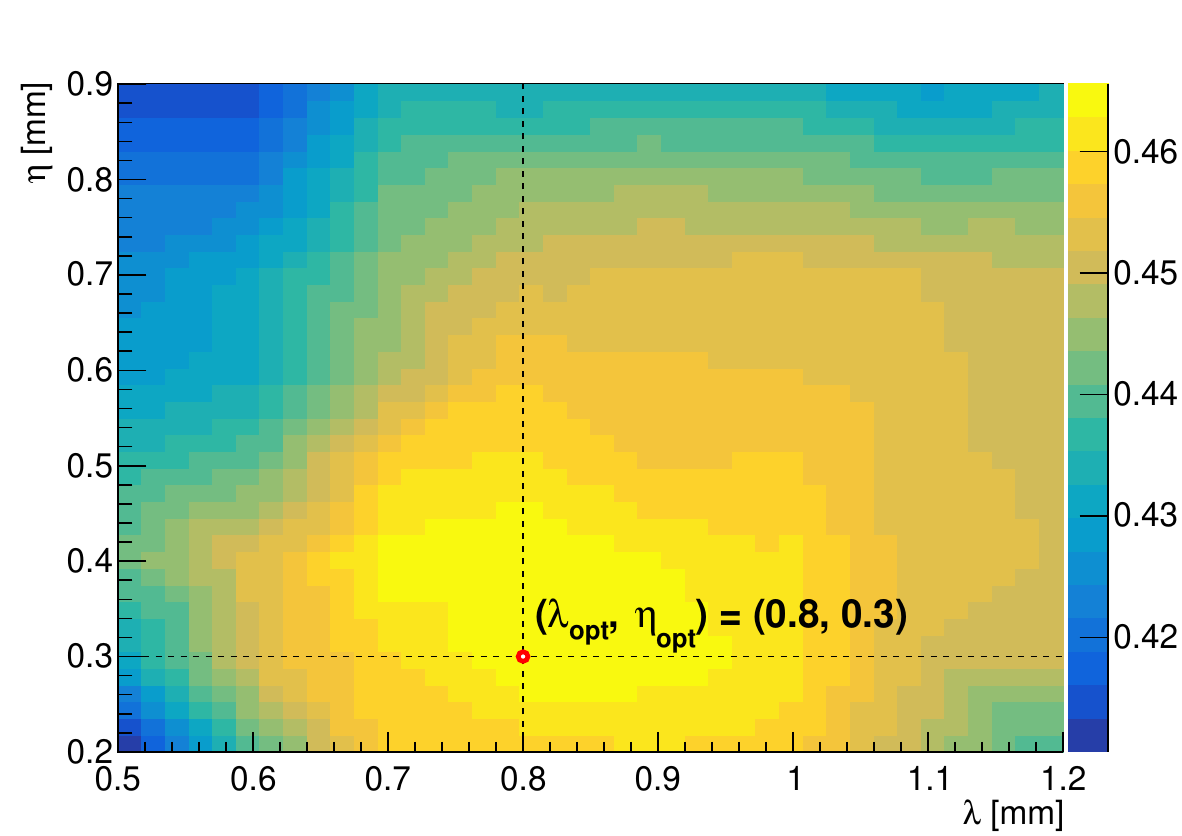}
	\caption{Turn reconstruction efficiency of the first stage as a function of $\lambda$ and $\eta$. The metric was measured after the self-refining.}
	\label{fig:twoLambda}		
	\centering
	\includegraphics[width=0.808\linewidth]{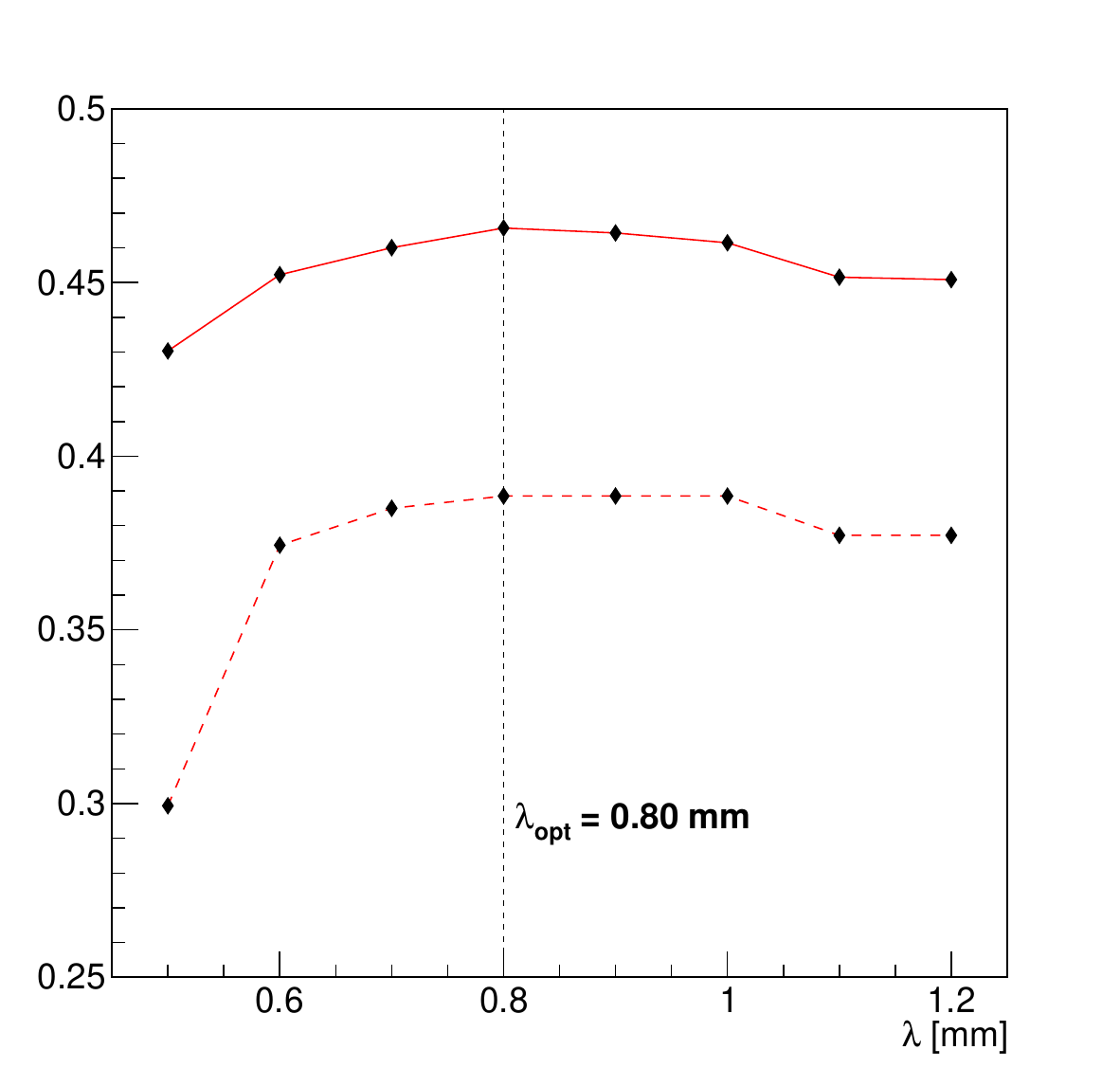}
	\caption{Turn reconstruction efficiency of the first stage with respect to $\lambda$ where $\eta$ is fixed to 0.3 mm. The dashed and solid lines represent the values before and after the self-refining, respectively.}
	\label{fig:reconfeff_vs_lambda}
	\includegraphics[width=0.808\linewidth]{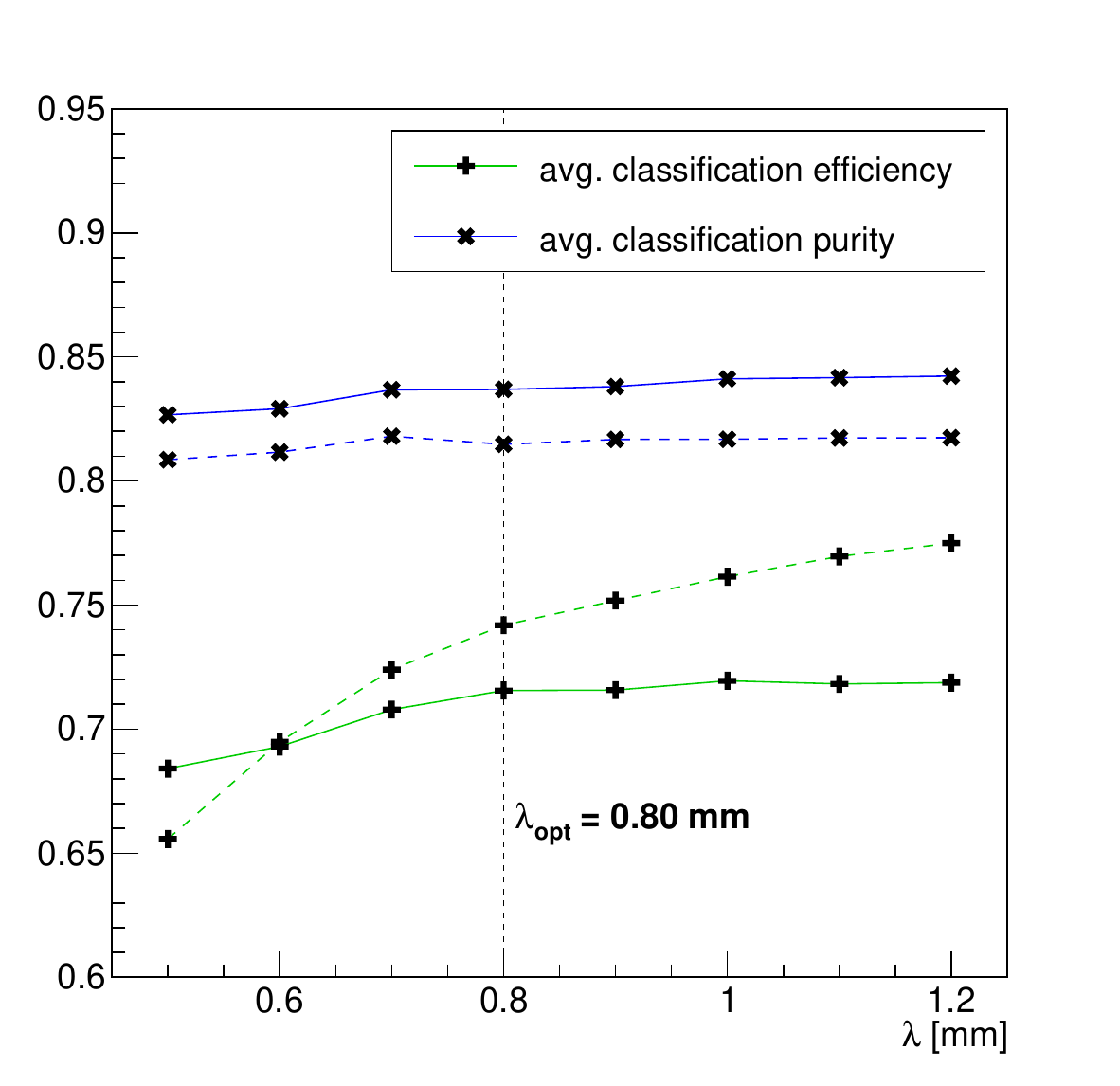}
	\caption{Averaged hit classification efficiency (green), and  hit classification purity (blue) of the first stage with respect to $\lambda$ where $\eta$ is fixed to 0.3 mm. The dashed and solid lines represent the values before and after the self-refining, respectively.}
	\label{fig:eff_pur_vs_lambda}
\end{figure}

\begin{figure}[t!]	
	\includegraphics[width=\lw\linewidth]{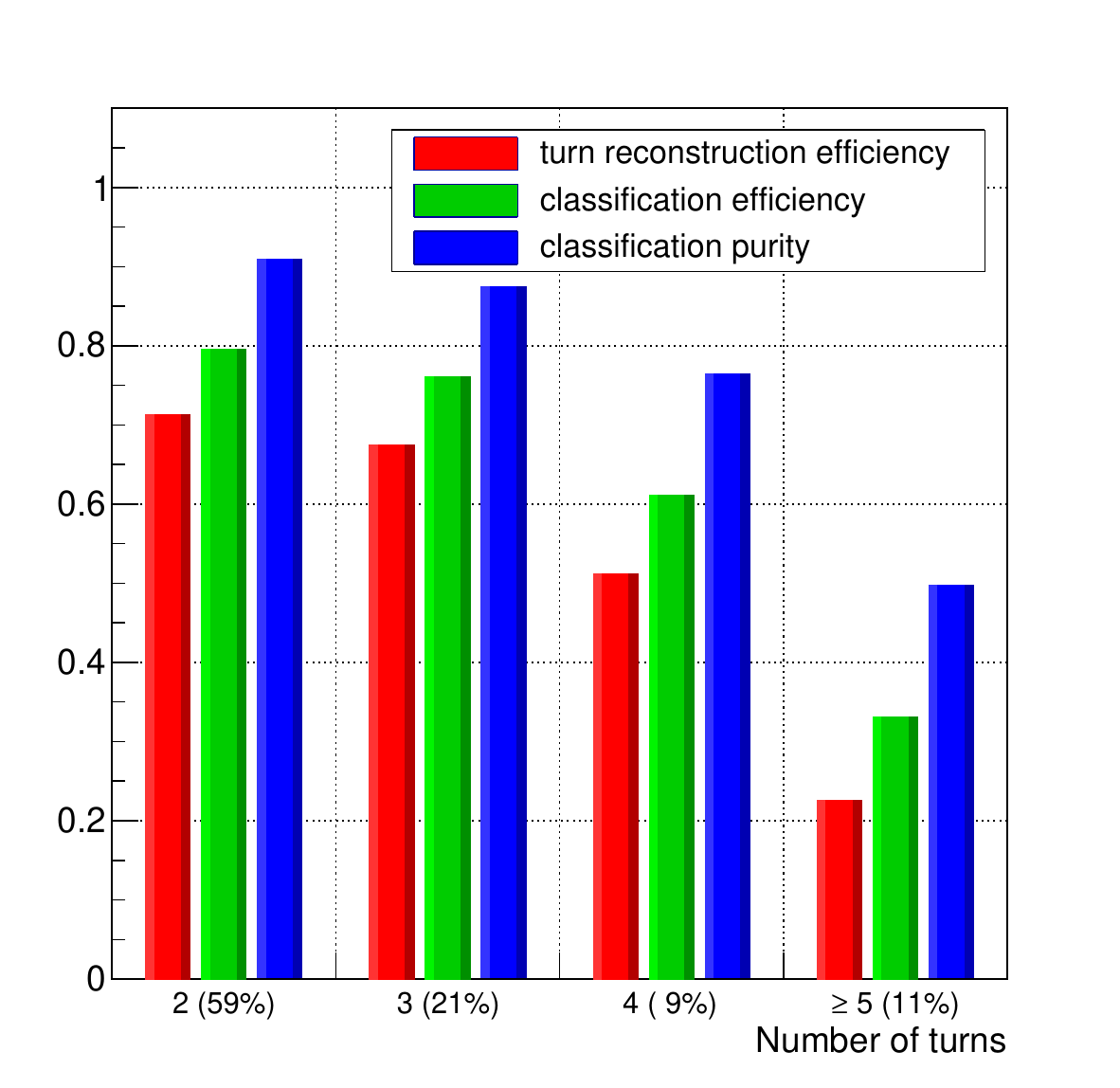}
	\caption{The bar chart of the turn reconstruction efficiency (red), the averaged hit classification efficiency (green), and the averaged hit classification purity (blue) of the first stage of event reconstruction after the self-refining vs. the number of turns of events. $\lambda$ was set to 0.8 $\textrm{mm}$, the optimized value. The percentages in parentheses represent the fraction of each event type.}
	\label{fig:eff_pur_vs_TN}
\end{figure}

\subsection{Metrics definitions}
\subsubsection{Hit classification metrics}
After every stage, the set of classified hits was divided into the subsets based on the turn index. The set of classified hits was considered representing the turn index of the largest subset. A \tit{hit classification efficiency} of the turn was defined as the ratio of the size of the largest subset to the actual number of hits induced by the represented turn partition. A \tit{hit classification purity} of the turn was defined as the ratio of the size of the largest subset to the size of the set of classified hits.
\subsubsection{Reconstruction efficiency metrics}
After applying the track fitting method, two metrics were defined to estimate the efficiency: (1) a \tit{turn reconstruction efficiency} and (2) an \tit{event reconstruction efficiency}. These metrics are for a turn partition and an event, respectively. The turn reconstruction efficiency was defined as the fraction of the events in which the turn fitted with the hits classified at the first stage satisfies the following track selection criteria: $\textrm{NDF}\geq \trm{max}(50,4\times N_L)$ and the reduced chi-square $(\chi^2/\textrm{NDF}) < 2$, where $N_L$ is the number of layers that have classified hits. The event reconstruction efficiency was defined as the fraction of the events that pass the following event selection criteria: any of the fitted tracks satisfies the track selection criteria.

\subsection{Cutoff values optimization}\label{sec:optimization}
For a cutoff value optimization, the signal electrons simulated by the Geant4 package \cite{GEANT4} were used. The drift distances of hits were smeared by a Gaussian distribution with a 0.15 mm standard deviation, corresponding to the measurement uncertainty \cite{COMET_TDR}. To reject events in undesirable shapes, we applied event preselection cuts in which electrons should pass at least five layers and the number of fired wires should be less than 250. This excludes 3\% of the multiple turn signal electrons that trigger the hodoscopes. 
\\
\indent 
The objective of the optimization process was to maximize the turn reconstruction efficiency at the first stage of the event reconstruction process. This is because the more precisely the track of the first stage is fitted, the better resolution we get for the longitudinal seeds of the next stage. Figure \ref{fig:twoLambda} shows how the two cutoff values ($\lambda, \eta$) affect the turn reconstruction efficiency after the self-refining. The tested ranges of $\lambda$ and $\eta$ are [0.5, 1.2 mm] and [0.2, 0.9 mm] with a 0.1 mm interval. The turn reconstruction efficiency was maximized when $\lambda$ and $\eta$ was 0.8 and 0.3 mm, respectively. \\
\indent
Figure \ref{fig:reconfeff_vs_lambda} shows the turn reconstruction efficiency as a function of $\lambda$ when $\eta$ is fixed to the optimized value. The metric was taken before and after the self-refining, and the improvement from the self-refining is discernible. The hit classification efficiency and purity averaged across the preselected events were also measured in the same condition, as shown in Fig. \ref{fig:eff_pur_vs_lambda}. After the self-refining, the purity increases whereas the efficiency drops. The higher purity, the less hits from other turn partitions are stolen during the first stage. However, there is also a downside from the lower efficiency where the unclassified hits of the first turn partition may hinder the hit classification of the next stages. \\
\indent The metrics from the optimized $\lambda$ and $\eta$ were broken down in terms of the turn numbers of the events, as shown in Fig. \ref{fig:eff_pur_vs_TN}. The performance improves with fewer turns since the hit overlaps are less significant.

\begin{figure}[!h]
	\centering
	\includegraphics[width=\lw\linewidth]{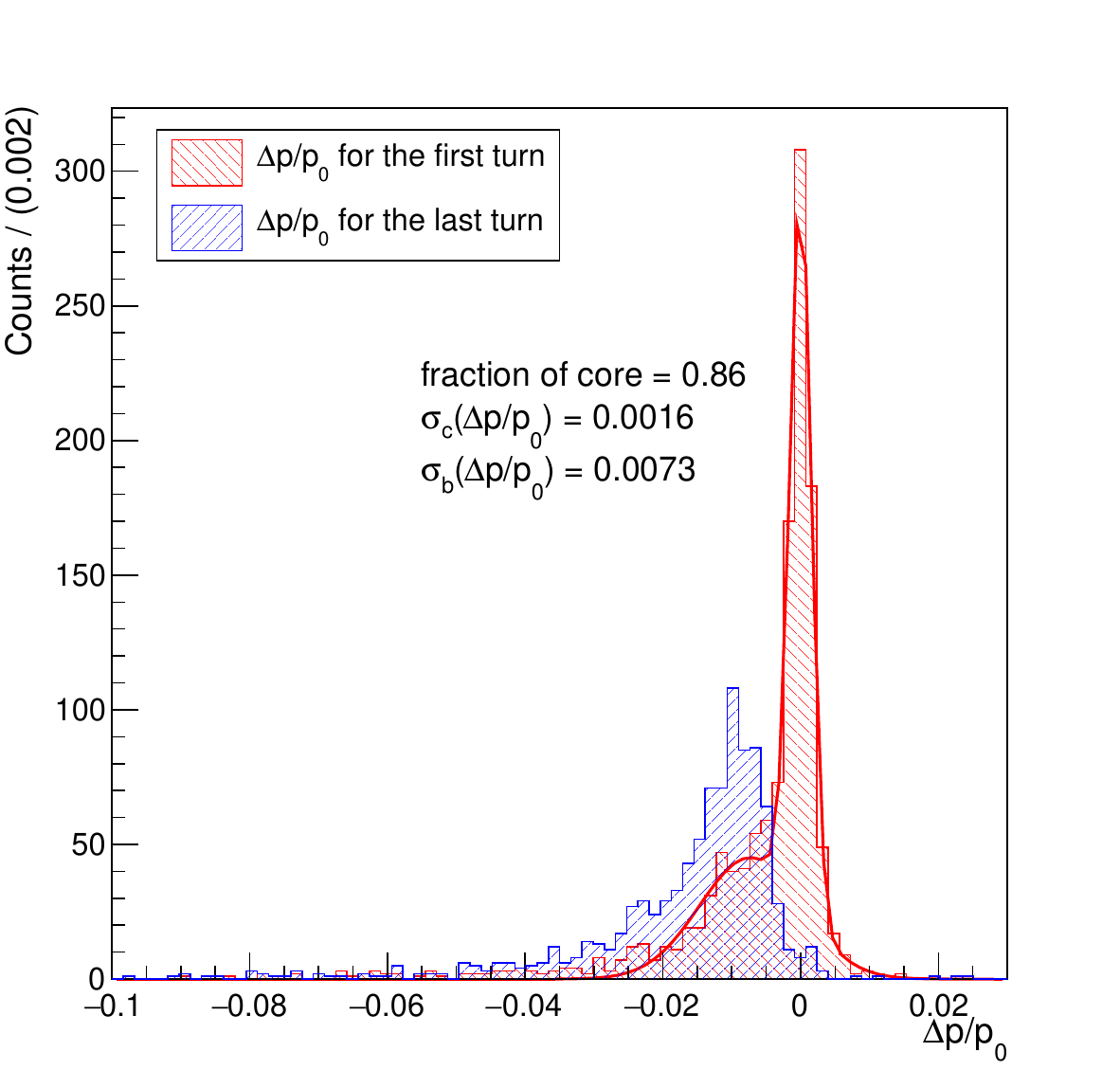}
	\caption{The histogram of $\Delta p/p_0$ for the first turn (red) and the last turn (blue) in the case of the signal electrons with multiple turns. The red line is the double Gaussian fitting function, where $\sigma_c$ and $\sigma_b$ represent the standard deviations of core and base part, respectively.}
	\label{fig:signal_residual}		
	\includegraphics[width=\lw\linewidth]{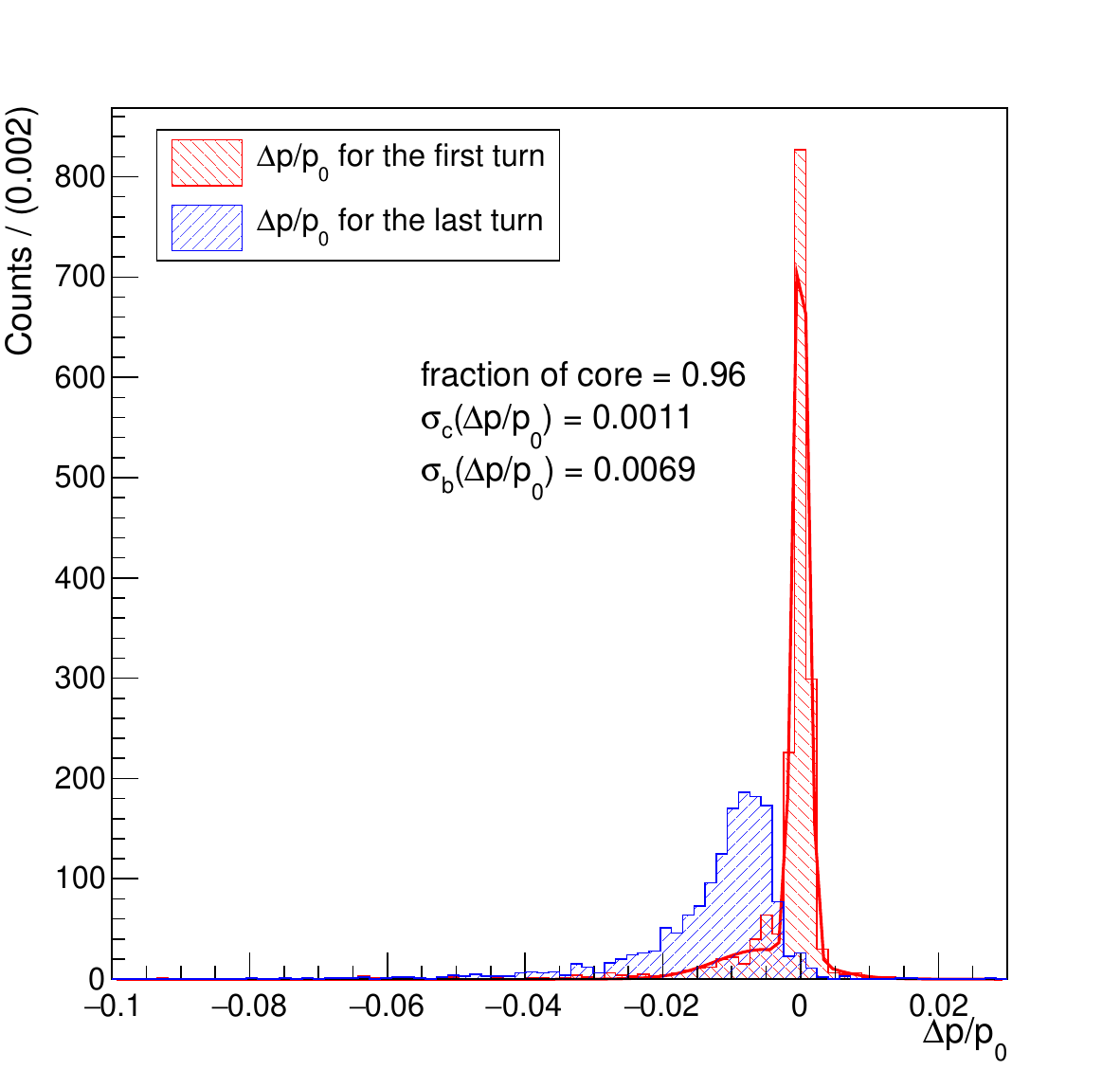}	
	\caption{The histogram of $\Delta p/p_0$ similar with Fig. \ref{fig:signal_residual} but in the case the signal electrons were simulated in the geometry without the wires.}
	\label{fig:signal_residual_nowire}			
\end{figure}

\begin{table}[t]
	\centering
	\caption{The comparisons of momentum resolution between the expected values from Eq. (\ref{eq:mom_res}) and the fitted values in Fig. \ref{fig:signal_residual} and \ref{fig:signal_residual_nowire}}
	
	\begin{tabular}{r r r r}
		\hline
		\hline
		& w/ wires  & w/o wires & comment \Tstrut  \\		
		\hline		                 
		$\sigma ( \Delta p /p_0 )$    & 0.0020 & 0.0011 & expected \Tstrut \Bstrut \\
		$\sigma_c ( \Delta p /p_0 )$  & 0.0016 & 0.0011 & measured \Tstrut \Bstrut \\        
		$\sigma_b ( \Delta p /p_0 )$  & 0.0073 & 0.0069 & measured \Tstrut \Bstrut \\                
		fraction of core              & 0.86   & 0.96 & measured \Tstrut \Bstrut \\
				
		\hline
		\hline
	\end{tabular}
	\label{tab:mom_resolution}
\end{table}

\begin{figure}[!h]
	\centering
	\includegraphics[width=\lw\linewidth]{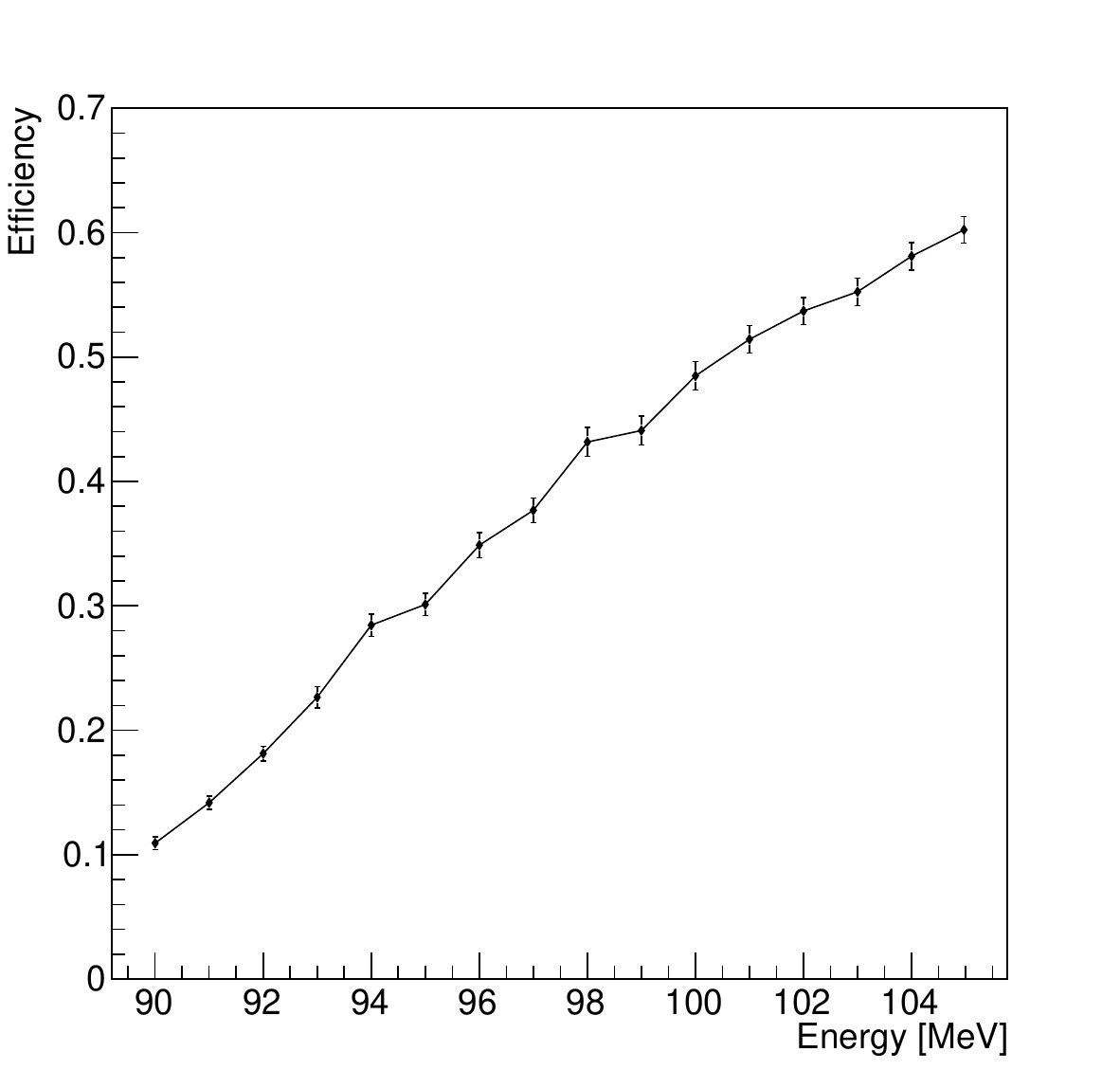}
	\caption{Event reconstruction efficiency of the multiple turn events as a function of the initial energy of electrons. The error bars were calculated assuming a binomial distribution, where the probability is the event reconstruction efficiency.}
	\label{fig:total_eff_vs_energy}
	
	\vspace{0.3 cm}
	
	\includegraphics[width=\lw\linewidth]{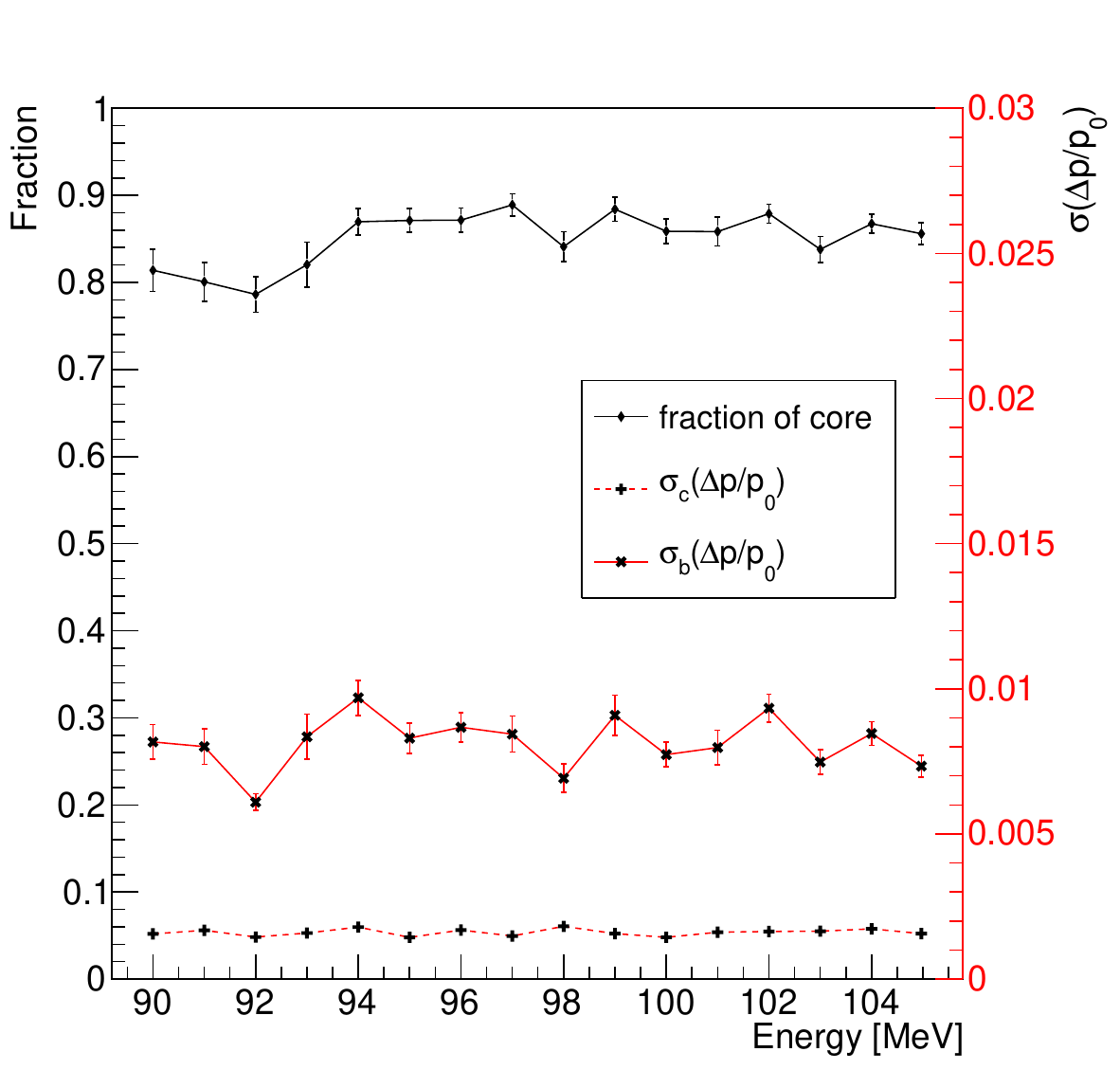}
	\caption{Double Gaussian fitting parameters on $\Delta p/p_0$ as a function of the initial energy of electrons. Left axis (black) is for the fraction of core part, and right axis (red) is for the standard deviations of core and base parts. The error bars were obtained from fitting errors.}
	\label{fig:fit_info}
	
\end{figure}

\subsection{Momentum resolution}
After using the optimized cutoff value, the momentum resolution of the multiple turn  signal electrons that passed the event selection criteria was investigated. The fitted tracks of an event were reordered based on their longitudinal distances to the triggered end of solenoid: the farthest and closest tracks were considered the first and last turn partitions, respectively. For the momentum resolution, a normalized momentum residual $(\Delta p/p_0)$ was defined, where $\Delta p$ is calculated by subtracting the true CDC entrance momentum of the first turn partition $(p_{0})$ from the fitted momentum of track $(p_{\textrm{fit}})$. In Fig. \ref{fig:signal_residual}, the distributions of $\Delta p/p_0$ are presented for the first and last turn partitions of signal electrons, respectively. Whereas the distribution of the last turn partition is shifted to the negative side due to the energy loss, the distribution of the first turn is fitted well with a double Gaussian function consisting of a core and base part. 
\\
\indent The momentum uncertainty of a detector can be predicted when the error of the track curvature $(\rho)$ is known \cite{KARIMAKI}: 
\begin{equation}\label{eq:mom_res}
\sigma \left( \frac{\Delta p}{p_0} \right) \approx \frac{\sigma (p_T)}{p_T} = \frac{\sigma (\rho) [\trm{mm}^{-1}]}{0.3B \trm{[T]}}p_T [\trm{MeV}/c],
\end{equation}  
where $\sigma$ with the parenthesis represents the uncertainty of the argument. $p$ and $p_T$ were approximated to be the same because mean $p_T$ of the multiple turn signal electrons is 101 MeV/c. $\sigma^2(\rho)$ can be represented by the square sum of two components \cite{GLUCKSTERN}:
\begin{equation}
\sigma^2(\rho) = [\sigma(\rho)_{\trm{res}}]^2 + [\sigma(\rho)_{\trm{ms}}]^2,
\end{equation}
where $\sigma(\rho)_{\trm{res}}$ and $\sigma(\rho)_{\trm{ms}}$ are from the spatial resolution of the drift distance measurement and from the multiple scattering in the detector, respectively. \\
\indent
$\sigma(\rho)_{\trm{res}}$ can be approximated under the assumptions of the homogeneous magnetic field and uniformly spaced measurements:
\begin{equation}\label{eq:curv_res}
\sigma(\rho)_{\trm{res}} \approx \frac{\epsilon}{L'^2}\sqrt{\frac{720}{N_{m}+4}},
\end{equation}
where $\epsilon$ is the resolution of the drift distance which was $150$ $\mu m$ for the simulated events, $L'$ is the length of the track projected on the transverse plane, and $N_m$ is the number of the measurements. If an electron track is generated from the center of the stopping target and $N_m$ is 50 for a turn partition, $\sigma(\rho)_{\trm{res}}$ becomes $1.4 \times 10^{-7}$ $\trm{mm}^{-1}$.
\\ \indent 
$\sigma(\rho)_{\trm{ms}}$ can be estimated as follows:
\begin{equation}
\sigma(\rho)_{\trm{ms}} \approx \frac{16 (\trm{MeV}/c)z}{L p \beta \cos^2{\lambda}} \sqrt{\frac{L}{X_0}},
\end{equation}
where $z$ is the charge of the particle in units of $e$, $L$ is the length of the track, $\beta$ is $v/c$, $\lambda$ is the pitch angle (i.e., $p\cos{\lambda}=p_T$), and $X_0$ is the radiation length of the detector material. The radiation length of the CDC gas mixture is $1.4 \times 10^{6} \trm{ mm}$, however, the effect from the cathode wires with shorter radiation length of 89 mm is not negligible. The effective radiation length of the composite materials can be obtained by averaging the density of the wires over the CDC volume \cite{RadiationLength}, where $X_0$ becomes $3.8 \times 10^{5} \trm{ mm}$. Under the same assumption used for Eq. (\ref{eq:curv_res}), $\sigma(\rho)_{\trm{ms}}$ was estimated to be $3.1 \times 10^{-6}$ $\trm{mm}^{-1}$ without considering the wires and $6.0 \times 10^{-6}$ $\trm{mm}^{-1}$ with the effective radiation length, respectively. $\sigma(\rho)_{\trm{res}}$ is negligible in either case. $\sigma(\Delta p/p_0)$ calculated from Eq. (\ref{eq:mom_res}) was 0.0020 with the effective radiation length with wires and 0.0011 when ignoring wires. \\ 
\indent Since the measured momentum resolution of the simulated events is supposed to be matched to the expected value in the case of no wires, for validation purpose, the event reconstruction was carried out for the events simulated with the geometry where the material of the wires is replaced by the gas mixture. Its distribution of $\Delta p/p_0$ is shown in Fig. \ref{fig:signal_residual_nowire} for comparison with the standard case of Fig. \ref{fig:signal_residual}. Table \ref{tab:mom_resolution} shows the expected momentum resolution and double Gaussian fitting parameters of both cases. We can see a good agreement between the expected momentum resolution and the standard deviation of core Gaussian $(\sigma_c (\Delta p /p_0 ))$ when the wire geometry is ignored. Although there is larger deviation in the momentum resolution for the standard case, this difference can be acceptable since the scattering at wires was approximated by using an effective radiation length. The base Gaussian, which is shifted to the negative side, was caused by the samples whose event reconstruction could not reconstruct the first turn partition but stopped at the earlier stages.

\subsection{Results for various electron energy} 
The event reconstruction efficiency was measured for various electron energy from 90 MeV to 104.97 MeV, as shown in Fig. \ref{fig:total_eff_vs_energy}. The efficiency of the signal electrons was about 60\%, which is higher than the first turn reconstruction efficiency measured in Fig. \ref{fig:twoLambda}. It is because, even though the criteria for the first turn reconstruction is not satisfied, those for later turns can be fulfilled. It was observed that the efficiency falls off as electron energy goes down. The electrons with lower energy usually leave fewer hits in the detector with lower transverse momentum, and this leads to a small NDF that is insufficient to pass the track selection criteria.
\\ 
\indent To verify consistency in the momentum resolution, the double Gaussian fitting parameters of $\Delta p/p_0$ distributions were extracted from each energy, as shown in Fig. \ref{fig:fit_info}. Despite some fluctuations in the standard deviations and fractions of the base parts, good agreement  among the distributions can be seen. 

\section{GPU-acceleration performance}\label{sec:results}

\subsection{Validation of GPU computation}
Having adopted two computing platforms for the event reconstruction, it is necessary to check whether their precisions are identical or within reasonable tolerance.  Some floating point operations conducted by GPU and CPU may not have the same results due to different standards in rounding \cite{GPUFLOP}. The rearranged algorithms in each platform also make the rounding occur in different orders. It should be noted that the GPU computation used here may not be deterministic for every runtime due to the atomic operations in which the order of additions is undetermined. However, the CPU computation results must be deterministic not only for every runtime but also for an arbitrary number of threads since the order of algorithm never changes during the seed scanning. \\
\indent The following parameters were compared between a GPU and CPU: (1) the RSS obtained at the first iteration of the first stage of event reconstruction and (2) the fitted momentum of the events that pass the event selection criteria in both platforms. It turned out that the difference in the RSS normalized by the CPU values is less than $O(10^{-12})$ which gives exactly equal results for the hit classification and fitted momentum.    

\begin{figure}[!h]
	\centering
	\includegraphics[width=\lw\linewidth]{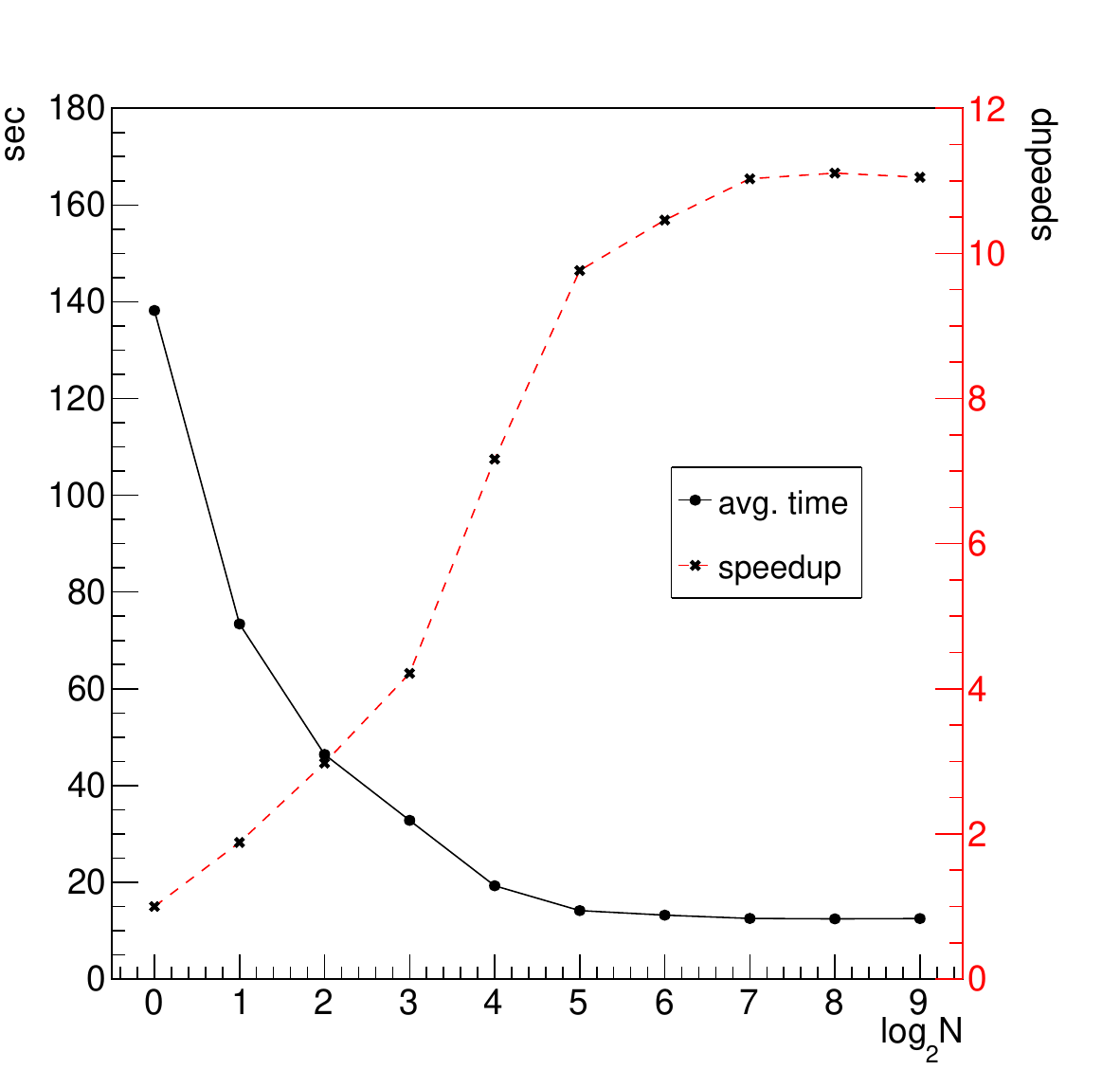}
	
	\caption{The graph of averaged computing time and speedup vs. the number of the CPU threads $(N)$, where $N=32$ $(\log_{2}N = 5)$ corresponds to the number of physical cores. The time was taken for the first stage of the seed scanning. The speedups are normalized for a single thread performance.}
	\label{fig:nThreads_cputime}
\end{figure}

\begin{table*}[h!]
	\centering
	\caption{Specifications of the tested GPU devices}
	\begin{tabular}{ r r r r r r r r r}
		\hline
		\hline
		\mr{2}{*}{Model}  & \mc{1}{l}{L1}    & \mc{1}{l}{L2}    & \mc{1}{l}{DRAM}   & \mc{1}{l}{DRAM}       & \mc{1}{l}{SMP}   & \mc{1}{l}{Number} & \mc{1}{l}{Double precision} & \mc{1}{l}{Launch}  \Tstrut \Bstrut \\
		                   & \mc{1}{l}{cache} & \mc{1}{l}{cache} & \mc{1}{l}{memory} & \mc{1}{l}{bandwidth}  & \mc{1}{l}{clock} & \mc{1}{l}{of SMP} & \mc{1}{l}{flop rate} & \mc{1}{l}{price}   \\
		\hline
		Tesla K40m    & 16 KiB & 1.5 MiB & 11.2 GiB & 288.4 GB/s & 745 MHz  & 15 & 1.43  TFLOPS & 7699 \$  \Tstrut \Bstrut \\ 
		Tesla K80     & 16 KiB & 1.5 MiB & 11.2 GiB & 240.5 GB/s & 824 MHz  & 13 & 1.37  TFLOPS & 5000 \$ \Bstrut \\ 
		Quadro P4000  & 48 KiB & 2.0 MiB & 7.93 GiB & 243.3 GB/s & 1.48 GHz & 14 & 0.166 TFLOPS & 815 \$ \Bstrut \\ 
		GTX 1070      & 48 KiB & 2.0 MiB & 7.93 GiB & 256.3 GB/s & 1.76 GHz & 15 & 0.211 TFLOPS & 379 \$ \Bstrut  \\ 
		\hline
		\hline
	\end{tabular}
	\label{tab:GPUspecs}
\end{table*}

\begin{table}[t]
	\centering
	\caption{Speedups of GPU devices over CPUs (two of E5-2630V3, 32 physical cores in total) with $N$ threads. $N=1$ is for the single thread performance, $N=32$ is for the number of the physical cores, and $N=128$ is for the optimal number of the threads, respectively.}
	
	\begin{tabular}{r r r r}
		\hline
		\hline
		\mr{2}{*}{Model}	  & \mc{3}{r}{Speedups over $N$ CPU threads} \Tstrut  \\		
		\cline{2-4}
		& $N=1$ & $N=32$ & $N=128$   \Tstrut  \\ 
		\hline
		K40m            & 203   & 21.1  & 18.5  \Tstrut \Bstrut \\		 
		K80             & 193   & 20.0  & 17.5 \Bstrut \\
		P4000           & 93    & 9.7   & 8.5  \Bstrut \\
		1070            & 120   & 12.5  & 10.9  \Bstrut \\						
		\hline
		\hline
	\end{tabular}
	\label{tab:GPUspeedups}
\end{table}

\begin{figure}[h]
	\centering
	\includegraphics[width=0.89\linewidth]{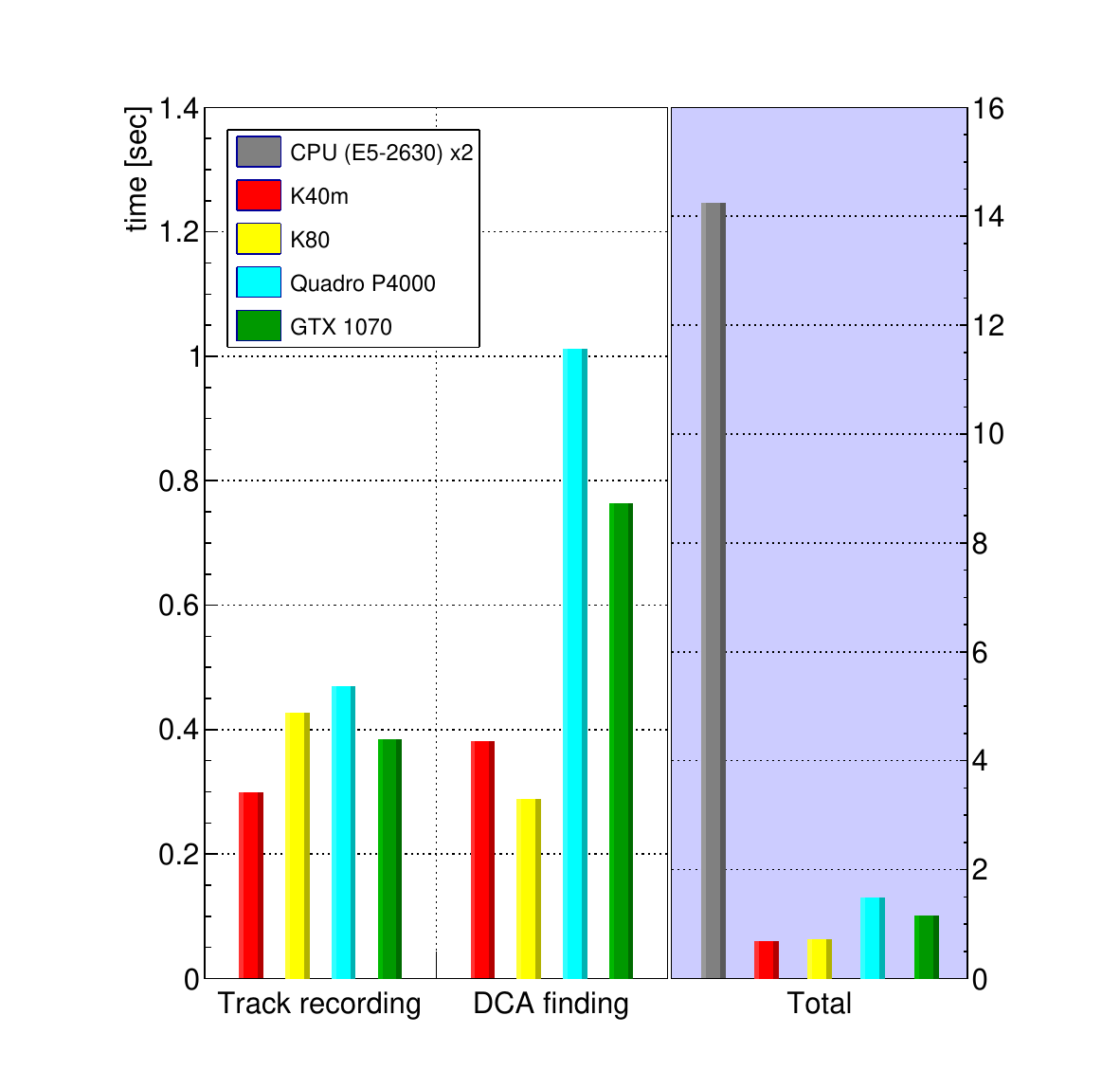}
	\caption{Averaged computing time of each GPU and CPUs (two of E5-2630V3) with 32 threads. The time was taken for the first stage of the seed scanning. The left part and the shaded right part show the time spent for each kernel function and their sums, respectively.}
	\label{fig:time_for_kernel}
\end{figure}

\begin{figure}[h!]
	\centering
	\includegraphics[width=0.89\linewidth]{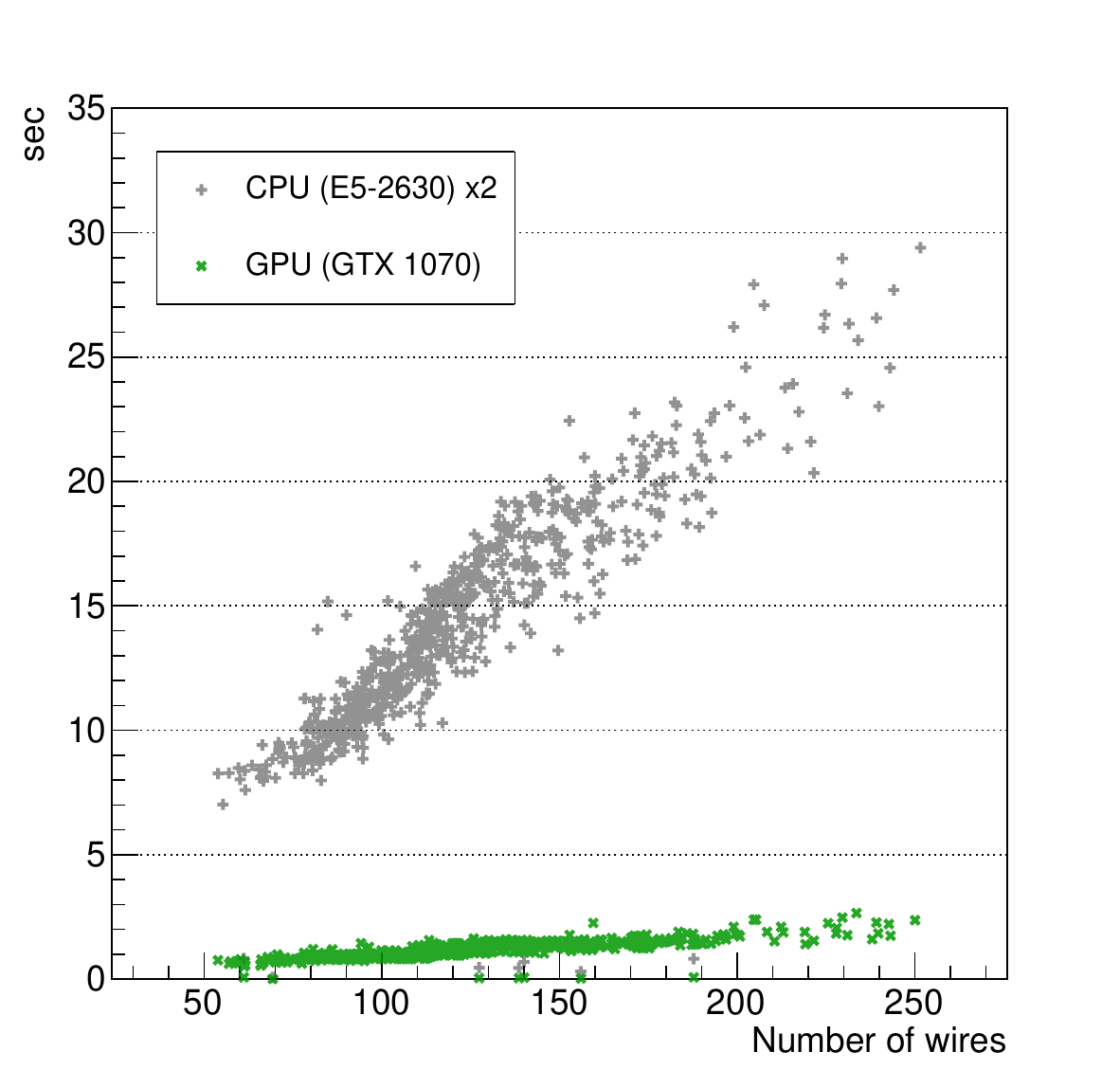}
	\caption{Computing time of the GPU (GTX 1070) and CPUs (two of E5-2630V3, 32 threads) with respect to the number of the wires in an event. The time was taken for the first stage of the seed scanning.}
	\label{fig:gpu_time_2D}
\end{figure}

\begin{table*}[t]
	\centering
	\caption{CUDA kernel metrics measured for each device. The compute and memory utilization levels are scaled from 0 to 10. All devices have the same values for the data load and store efficiency.}
	\begin{tabular}{l l r  r r r r}
		\hline
		\hline
		\mr{2}{*}{Kernel} & \mr{2}{*}{Model} & \mc{1}{l}{Compute} & \mc{1}{l}{DRAM} & \mc{1}{l}{L2 cache} & \mc{1}{l}{Data load} & \mc{1}{l}{Data store} \Tstrut  \\ 
		                  &                  & utilization        & utilization  & \mc{1}{l}{missing rate}    &  \mc{1}{l}{efficiency} & \mc{1}{l}{efficiency}  \\
        \hline	                               
                          &  K40m   & 2    & 4  & 47.6\%   & \mr{4}{*}{25.8\%} & \mr{4}{*}{25.5\%} \Tstrut \Bstrut \\
         \ttt{Track}      &  K80    & 3    & 5  & 50.6\%  &  &  \Bstrut \\
         \ttt{recording}  &  P4000  & 10   & 2  & 12.5\%  &  &   \Bstrut \\
                          &  1070   & 10   & 1  & 12.4\%  &  &   \Bstrut \\
         \hline
                          &  K40m   & 5    & 3  & 56.7\%  & \mr{4}{*}{25.4\%} & \mr{4}{*}{25.0\%} \Tstrut \Bstrut \\
        \ttt{DCA}         &  K80    & 5    & 2  & 57.7\%  &  &   \Bstrut \\
        \ttt{finding}	  &  P4000  & 10   & 1  & 0.91\%  &  &   \Bstrut \\
                          &  1070   & 10   & 1  & 0.97\%  &  &   \Bstrut \\      
		\hline
		\hline
	\end{tabular}
	\label{tab:GPUmetrics}
\end{table*}

\subsection{Multi-threaded CPU speedup}
\indent The speedup of a multi-threaded CPU  was investigated for $2^n$ $(n=0, 1, \ldots, 9)$ threads . Figure \ref{fig:nThreads_cputime} shows an averaged computing time and speedups over the single thread performance where the time was measured for the first stage of the seed scanning. Since two of the CPUs (Intel Xeon E5-2630V3, 2.4 GHz, 16 physical cores each) in a single node ran in parallel with 32 physical cores, the speedup increased drastically up to 32 threads. The tendency of speedup was not linear below 32 threads because the cache shared by the cores easily becomes saturated with the workloads of few threads. Conversely, the speedup slightly increased above 32 threads since the jobs are scheduled more efficiently to make the cores less idle.

\subsection{GPU speedup}
\indent 
The major specifications of the tested GPU models are listed in Table \ref{tab:GPUspecs}. The K80 contains two devices and its specifications and benchmarked parameters are from a single device. A DRAM is the off-chip device memory that exchanges data with the host memory and caches. An L1 and L2 cache are on-chip memory for fast memory accesses. Every SMP is served by different L1 caches whereas there exists only one L2 cache shared by all SMPs. For the benchmarked models, an actual caching of data happens only in the L2 cache unless a particular option is given to the compiler. If all resources in the L2 cache are running out, the data is evicted to the DRAM that has the slowest access speed. \\
\indent 
The Tesla K40m and K80 have slower clock rates and less resources in caches and registers. However, their double precision computing units  have extraordinary efficiency that enables them to have much higher flop rates. The computing speed may scale linearly with the flop rate in an ideal case where the algorithm is compute-bound.
\\
\indent Figure \ref{fig:time_for_kernel} shows the computing time spent during the first stage of the seed scanning. GPU speedups over the multi-threaded CPU are listed in Table \ref{tab:GPUspeedups}: The Tesla K40m showed the best performance in the speedup per device. Nevertheless, the GTX 1070 might be a better choice considering the price. The linearity of the computing time with respect to the number of wires was also confirmed as shown in Fig. \ref{fig:gpu_time_2D}. \\
\indent 
Although the Tesla products outperform others, their speedups are not proportional to the flop rates. This implies that the algorithms are not compute-bound and there might be some bottlenecks in memory traffic, which leaves computing units idle. Table \ref{tab:GPUmetrics} lists the kernel metrics measured by the profiler \cite{NVProf} to identify the stall reasons in the Tesla products. The measured compute utilization levels clarify that the algorithms are not compute-bound in the Tesla products. A data load/store efficiency is the ratio of the requested data transaction to the actual data transactions. Around 25\% of the efficiency is from a misaligned memory accessed pattern. The size of a double variable is 8 bytes and that of a data access pattern is 32 bytes; thus, one data request requires four transactions. An L2 cache missing rate represents the fraction of L2 cache data evicted to the DRAM. The smaller caches and registers of the Tesla products cause a higher missing rate, which increases the DRAM utilization level. In conclusion, the data clogging of Tesla products is the combined effect of the low data transaction efficiency and small memory resources.

\subsection{Estimation on offline analysis time}
The time required for offline analysis was estimated only for the DIO multiple turn events because the analysis time for the single turn events would be relatively small given that a simpler algorithm is applied. We also assumed that the contributions from other backgrounds are negligible. The number of DIO multiple turn events $(N_{\trm{DIO}})$ to be analyzed during the COMET Phase-I experiment was estimated as follows:
\begin{equation}\label{eq:NmuEstimation}
  N_{\trm{DIO}} = N_\mu \times f_{\trm{DIO}} \times P_{E>70} \times A_{\trm{trig}} \times A_{\trm{multi}} \times A_{\trm{cut}},
\end{equation}
where $N_\mu$ is the total number of stopped muons in the targets, $f_{\trm{DIO}}$ is the fraction of DIO, $P_{E>70}$ is the probability that the initial energy of DIO electron is higher than 70 MeV \cite{DIOspectrum}, $A_{\trm{trig}}$ is the ratio of triggering events, $A_{\trm{multi}}$ is the ratio of multiple turn events, and $A_{\trm{cut}}$ is the ratio of events that satisfy the preselection cuts introduced in Subsection \ref{sec:optimization}. The values of each parameter are listed in Table \ref{tab:NmuEstimation}, where $N_{\trm{DIO}}$ is estimated as $1.4\times 10^{7}$. In the case of the GTX 1070, the ratio of the seed scanning time to the wall time is about 15\%, and the rest was dominated by the self-refining process carried out by the CPU. The wall time per event is 8.1 seconds which requires around 1300 days of offline analysis time.

\begin{table}[t]
	\centering
	\caption{The values of the parameters required in Eq. (\ref{eq:NmuEstimation}), where $N_{\trm{DIO}}$ is estimated as $1.4\times 10^{7}$}	
	\begin{tabular}{r r r r r r}
		\hline
		\hline
		\mc{1}{l}{$N_\mu$} & \mc{1}{l}{$f_{\trm{DIO}}$} & \mc{1}{l}{$P_{E>70}$} & \mc{1}{l}{$A_{\trm{trig}}$} & \mc{1}{l}{$A_{\trm{multi}}$} & \mc{1}{l}{$A_{\trm{cut}}$}  \Tstrut \Bstrut \\
		\hline
		$1.5\times 10^{16}$ & 0.39 & $3.8 \times 10^{-6}$ & 0.016 & 0.47 & 0.08 \Tstrut \Bstrut \\
		\hline
		\hline
	\end{tabular}
	\label{tab:NmuEstimation}
\end{table}

\begin{figure}[t]
	\centering
	\includegraphics[width=\lw\linewidth]{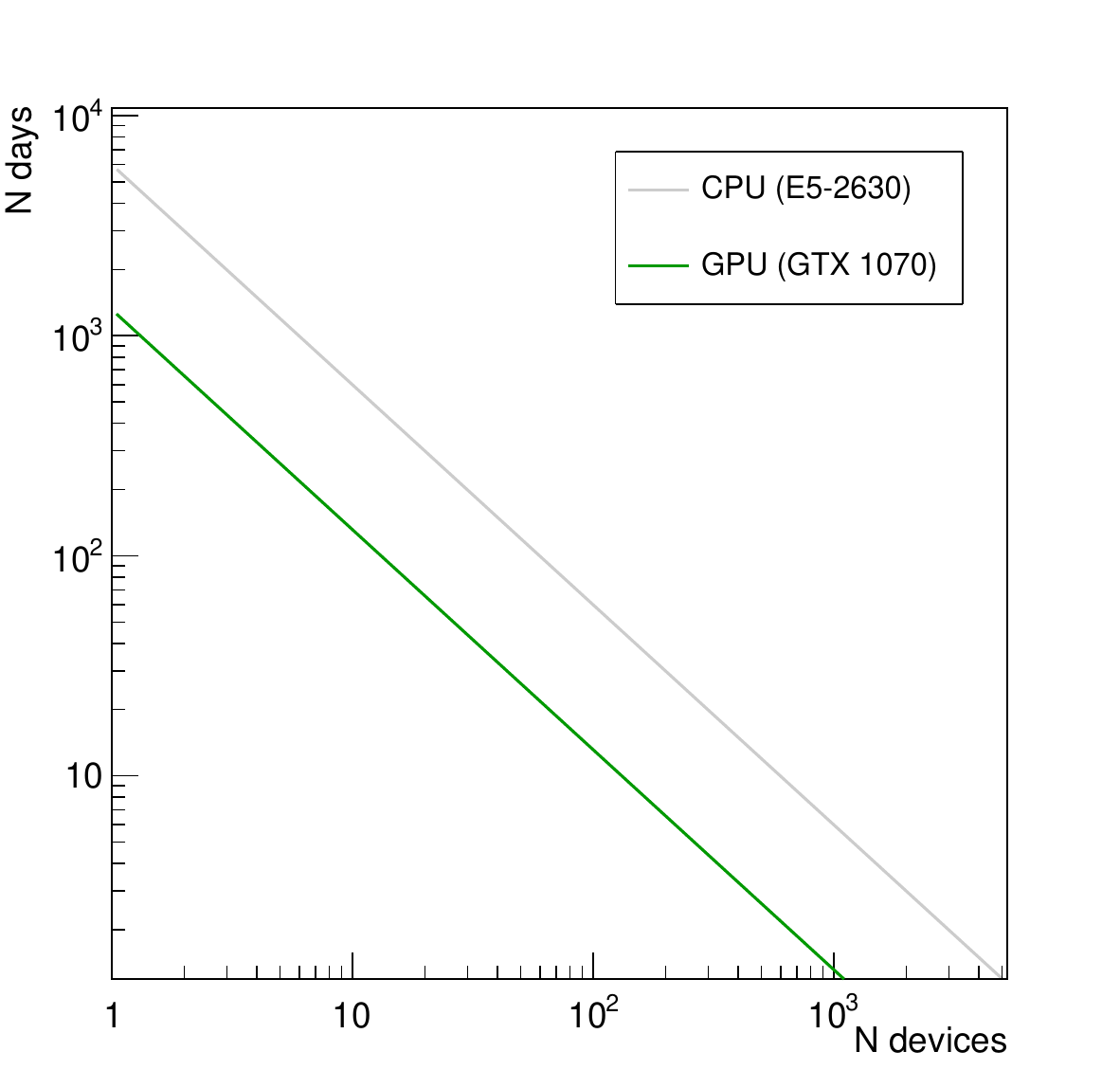}
	
	\caption{Log-log plot of the offline analysis time in day unit vs. the number of devices for each platform}
	\label{fig:nDevice_estimation}
\end{figure}

According to the GPU speedups of Table \ref{tab:GPUspeedups} and the ratio of the seed scanning to the wall time, we can naively estimate that one GTX 1070 can replace five E5-2630V3. Figure \ref{fig:nDevice_estimation} shows the estimation of the offline analysis time with respect to the number of devices. For example, 200 CPUs corresponding to 3200 cores are required to finish the analysis in a month while a GPU cluster requires 44 devices. Considering the launch prices and power consumption of the GTX 1070 (379 USD, 150 W) and E5-2630V3 (670 USD, 85 W), it is more economical to build up a cluster with the GPUs.

\section{Discussion}\label{sec:discussion}
\subsection{Discussion on the event reconstruction}
The hit-to-turn classification problem has been an obstacle to event reconstruction of multiple turn events in the COMET Phase-I experiment. To resolve it, the events were reconstructed turn-by-turn by finding the best seed of position and momentum for each turn and classifying the hits based on their distances to a track extrapolated from the seed. Since the constraints on longitudinal components were provided only for the last turn partition, the event reconstruction started from the end of the solenoid and continued in the backward direction. To reduce the total extrapolation length, the hit classification of a turn was done for two half-turns, each of which was extrapolated from the CDC entrance and exit, respectively. The complete sets of classified hits were obtained by combinatorially merging the half sets from each half turn, and the best set was selected based on the track fitting quality. The set of hits was self-refined by applying the Kalman filtering iteratively to improve the hit classification performance. The longitudinal seeds of the other turn partitions were obtained by backward-extrapolating the lastly fitted turn partition.  Among the fitted tracks that passed the criteria, the track farthest from the triggering side was considered the first turn partition whose momentum was read for the event. \\  
\indent As a result, the reconstruction efficiency of multiple turn events of about 60\% was achieved with an acceptable momentum resolution, even though the momentum resolution was limited by the hit classification performance. The consistency in the momentum resolution across a wide range of electron momenta is a noticeable achievement showing that the algorithm works not only for the signal but also for the DIO electrons with lower momenta. \\
\indent The performance of hit classification is limited due to the random behaviors caused by multiple scattering and measurement uncertainty. Hence, the self-refining method was introduced for the track to exchange its hits with the unclassified hits. For further improvement, we can consider implementing a method where the track exchanges the hits used in other tracks to retrieve the stolen hits. \\
\indent The application of the algorithm will be extended to the byproducts from other muon decay as well as the electrons emitted from the stopping targets. They include (1) a 92.3 MeV positron emission from aluminum target through $\mu^- + N(A,Z) \rightarrow e^+ + N(A,Z-2)$ which is a lepton number violation process forbidden in the Standard Model \cite{BYEO}, and (2) an electron and positron emission from the pair production, followed by the radiative muon capture. The same algorithm with an opposite charge is also applicable to the positron events as we have shown that it works down to the 90 MeV electrons. Meanwhile, the reconstruction of the pair production events might require more dedicated study.

\subsection{Discussion on the GPU-acceleration}
\indent 
The algorithm for the hit classification and seed scanning was parallelized to gain high throughput by allocating the subtasks to each thread of a device. The CUDA code for GPUs was separated into two kernel functions to utilize device resources with better efficiency. On the other hand, the CPU codes were written as a single unified function to fit the algorithm into its device characteristics. The same internal RKN extrapolation codes were shared by both devices to increase the maintainability. The validation tests showed that the results from the CPU and GPU were equal to each other. Two multi-threaded CPUs, each of which has 16 cores, obtained an order of magnitude of speedup over a single CPU thread while the GTX 1070 of Pascal architecture pushed the speedup one order of magnitude further.\\  
\indent The offline analysis was estimated to be finished in a month with a few dozen GPU devices or a few hundred CPUs. Should an extension be required for a given offline analysis timeline, it should be decided whether to use the CPU or GPU based on the current resources and the expected gain. In case of building GPU clusters, the products with newer architectures such as Volta and Turing can be considered. An RTX series of Turing, which is an upgraded version of the GTX series, might be a promising option for future setups. 


\section*{Acknowledgements}
We thank Soohyung Lee for setting the GPUs in the server cluster. We also thank Jim Linnemann for helpful comments. The work of B.Y. and M.L. was supported by IBS-R017-D1-2020-a00 of the Institute for Basic Science, the Republic of Korea. The work of Y.K. was supported by the JSPS KAKENHI Grant No. 18H04231.

\bibliography{mybibfile}

\end{document}